\documentclass[11pt,notitlepage,preprintnumbers,longbibliography,nofootinbib]{revtex4-1}
\usepackage{mathdots,amsmath,amsfonts,amssymb,geometry}
\usepackage[utf8]{inputenc}
\usepackage{footmisc}
\usepackage{lmodern}
\usepackage{microtype}
\usepackage{multirow}
\usepackage{array,mathtools,booktabs}

\newcolumntype{C}{>{$}c<{$}}

\AtBeginDocument{
\heavyrulewidth=.08em
\lightrulewidth=.05em
\cmidrulewidth=.03em
\belowrulesep=.65ex
\belowbottomsep=0pt
\aboverulesep=.4ex
\abovetopsep=0pt
\cmidrulesep=\doublerulesep
\cmidrulekern=.5em
\defaultaddspace=.5em
}

\usepackage{mathrsfs}
\usepackage{bbm}
\usepackage{epsfig}
\usepackage{latexsym}
\usepackage{slashed}
\usepackage{xcolor}
\usepackage{amsthm}
\usepackage{amscd,bm}
\usepackage{amstext}

\usepackage{hyperref}
\hypersetup{
    colorlinks=true,
    linkcolor=blue,
    filecolor=blue,      
    urlcolor=blue,
    citecolor=blue,
}

\usepackage{graphicx}

\renewcommand{\arraystretch}{1.2}

\numberwithin{equation}{section}

\newcommand{\be}{\begin{equation}}
\newcommand{\ee}{\end{equation}}
\newcommand{\ba}{\begin{eqnarray}}
\newcommand{\ea}{\end{eqnarray}}
\newcommand{\bi}{\begin{itemize}}
\newcommand{\ei}{\end{itemize}}
\newcommand{\ben}{\begin{enumerate}}
\newcommand{\een}{\end{enumerate}}

\DeclareMathOperator{\arccoth}{arccoth}
\DeclareMathOperator{\arctanh}{arctanh}

\newcommand{\dd}{\mathrm{d}}

\DeclareMathOperator{\sign}{sign}

\begin{document}
\mbox{}
\vspace{8mm}
\title{\Large{Second order equilibrium transport in strongly coupled $\mathcal{N} = 4$ supersymmetric $SU(N_c)$ Yang-Mills plasma via holography}}

\author{Sebastian Grieninger}
\email{sebastian.grieninger@gmail.com}
\affiliation{Instituto de Fisica Teorica UAM/CSIC and Departamento de Fisica Teorica, Universidad Autonoma de Madrid, Campus de Cantoblanco, ES-28049 Madrid, Spain}

\author{Ashish Shukla}
\email{ashukla@perimeterinstitute.ca}
\affiliation{Perimeter Institute for Theoretical Physics, 31 Caroline Street North, Waterloo, Ontario N2L 2Y5, Canada}

\preprint{IFT-UAM/CSIC-21-56}

\begin{abstract}
\vspace{1cm}
\centering\begin{minipage}{\dimexpr\paperwidth-6cm}
    A relativistic fluid in 3+1 dimensions with a global $U(1)$ symmetry admits nine independent static susceptibilities at the second order in the hydrodynamic derivative expansion, which capture the response of the fluid in thermal equilibrium to the presence of external time-independent sources. Of these, seven are time-reversal $\mathbb{T}$ invariant and can be obtained from Kubo formulas involving equilibrium two-point functions of the energy-momentum tensor and the $U(1)$ current. Making use of the gauge/gravity duality along with the aforementioned Kubo formulas, we compute all seven $\mathbb{T}$ invariant second order susceptibilities for the $\mathcal{N} = 4$ supersymmetric $SU(N_c)$ Yang-Mills plasma in the limit of large $N_c$ and at strong 't-Hooft coupling $\lambda$. In particular, we consider the plasma to be charged under a $U(1)$ subgroup of the global $SU(4)$ R-symmetry of the theory. We present analytic expressions for three of the seven $\mathbb{T}$ invariant susceptibilities, while the remaining four are computed numerically. The dual gravitational description for the charged plasma in thermal equilibrium in the absence of background electric and magnetic fields is provided by the asymptotically AdS$_5$ Reissner-Nordstr\"{o}m black brane geometry. The susceptibilities are extracted by studying perturbations to the bulk geometry as well as to the bulk gauge field. We also present an estimate of the second order transport coefficient $\kappa$, which determines the response of the fluid to the presence of background curvature, for QCD, and compare it with previous determinations made using different techniques.
    \end{minipage}
\end{abstract}

\maketitle

\newpage

{\hypersetup{linkcolor=black} \tableofcontents}

\newpage
\section{Introduction and basic set-up}
\label{Introduction}
The holographic principle and its realization via the Anti de Sitter/Conformal Field Theory (AdS/CFT) correspondence \cite{Maldacena:1997re, Gubser:1998bc, Witten:1998qj} has been a powerful tool in exploring the properties of strongly coupled quantum field theories which are not amenable to a perturbative analysis. The correspondence allows one to study strongly interacting quantum systems via a dual gravitational description in an asymptotically AdS spacetime, and has been used extensively to explore several phenomena in high energy physics as well as condensed matter systems, including transport phenomena at finite temperature (see \cite{Aharony:1999ti, DHoker:2002nbb, Polchinski:2010hw, Hubeny:2014bla, Natsuume:2014sfa, ammon_erdmenger_2015, nastase_2015} for reviews, including some applications). 

In this article, we employ the AdS/CFT correspondence to study the equilibrium thermodynamic properties of a strongly coupled charged plasma at a finite temperature and chemical potential. The particular holographic quantum system we have in mind is the $\mathcal{N} = 4$ supersymmetric $SU(N_c)$ Yang-Mills plasma at finite temperature and chemical potential in 3+1 dimensions in thermal equilibrium. The theory enjoys a global $SU(4)$ R-charge symmetry, and we consider it to be charged under a $U(1)$ subgroup of the full R-symmetry group.\footnote{A comprehensive overview of R-symmetry charges and chemical potentials in $\mathcal{N} = 4$ SYM theory appears in \cite{Yamada:2006rx}.} In fact, this $U(1)$ is the diagonal $U(1)$ of the maximal Abelian subgroup $U(1)^3$ of $SU(4)$ \cite{Chamblin:1999tk}. An important point to keep in mind is that this $U(1)$ symmetry is axial in nature, implying that the associated chemical potential $\mu$ and current $J^\mu$ are axial too.\footnote{An axial chemical potential is parity odd but charge conjugation even.} Due to the chiral anomaly of the theory \cite{Witten:1998qj, Freedman:1998tz}, the current $J^\mu$ is not conserved, and its non-zero divergence is proportional to the product of external electric and magnetic fields applied to the system: $\nabla_\mu J^\mu \propto E\cdot B$. However, in the present article, we will study the transport properties of the plasma in the absence of any external electric and magnetic fields, thereby restoring current conservation.

In the limit of large $N_c$ and with the 't-Hooft  parameter $\lambda \equiv g_{\textrm{YM}}^2 N_c \gg 1$ \textit{i.e.}\ at strong coupling, the super-Yang-Mills plasma in thermal equilibrium, in the absence of background electric and magnetic fields, is holographically dual to a charged Reissner-Nordstr\"{o}m black brane solution of the five dimensional Einstein-Maxwell-Chern-Simons theory in asymptotically $\text{AdS}_5$ spacetime \cite{Chamblin:1999tk, Cvetic:1999ne}.\footnote{The dual bulk solution is different, and in general is to be constructed numerically, once the plasma is kept in external electromagnetic fields. See for instance \cite{Ammon:2017ded, Ammon:2020rvg} for recent discussions where charged magnetic black brane solutions have been constructed, dual to the charged $\mathcal{N} = 4$ super-Yang-Mills plasma kept in a background magnetic field, and the resulting transport properties of the plasma have been studied.} The transport  properties of the plasma can be extracted by studying perturbations to this black brane geometry. Studying transport in the super-Yang-Mills plasma provides an indirect route for a better understanding of transport in strongly coupled systems found in nature, such as the quark-gluon plasma (QGP). Produced in relativistic heavy ion collisions, the QGP is strongly coupled and can be considered conformal at the very high energy scales involved, since the quark masses can be neglected at such high energies to a very good approximation. As the strongly coupled $\mathcal{N} = 4$ super-Yang-Mills plasma can be probed via holography, wherein its properties can be studied using a dual gravitational description, it provides a simpler and tractable model for the much more complicated dynamics of the QGP which is hard to probe via direct computations \cite{Chesler:2015lsa}.\footnote{A simple holographic model for QCD wherein the breaking of conformality is captured by a non-trivial dilaton profile in the bulk, dual to a marginally relevant operator on the boundary, appears in \cite{Bigazzi:2010ku}.}

It has been observed experimentally that QGP behaves like a low viscosity relativistic fluid \cite{Romatschke:2017ejr}, with the shear viscosity to entropy density ratio close to the Kovtun-Son-Starinets bound $\eta/s \ge 1/4\pi$ \cite{PhysRevLett.94.111601}. Its dynamics can thus be modeled using the formalism of relativistic viscous hydrodynamics \cite{Baier:2007ix, Bhattacharyya:2008jc}. Hydrodynamics provides an effective description for the dynamics of many-body systems at or near thermal equilibrium, and is a vast subject in itself (see \cite{Romatschke:2009im, Kovtun:2012rj, Jeon:2015dfa, Romatschke:2017ejr} for pedagogical reviews). It owes its successes to the idea that the effective dynamics of such near-equilibrium systems can be captured in terms of a small number of variables associated to the system. These hydrodynamic variables are typically the local fluid four-velocity $u^\mu$ and the temperature $T$. In case the system also has a conserved $U(1)$ charge, then the associated chemical potential $\mu$ also becomes a hydrodynamic variable. The dynamics of the near-equilibrium system is governed by the conservation equation for the energy-momentum tensor $T^{\mu\nu}$, along with the conservation equation for the current $J^\mu$. The hydrodynamic approximation asserts that $T^{\mu\nu}$ and $J^\mu$ each admits a derivative expansion, with the transport parameters of the system, such as viscosities and conductivities, appearing as coefficients in this expansion. Note that the transport parameters are functions of the state of the system, parametrized by the temperature and chemical potential, and act as inputs to the macroscopic hydrodynamic description \textit{i.e.}\ they have to be computed using the underlying microscopic theory describing the system.

In the limit of thermal equilibrium, as was pointed out in \cite{Banerjee:2012iz, Jensen:2012jh}, the physical properties of the fluid can further be encoded in a static \textit{generating functional} $\mathrm{W}[g_{\mu\nu}, A_\mu]$. This generating functional is a functional of time-independent  external sources, which source the conserved currents of the system.\footnote{A more recent approach based on the Schwinger-Keldysh closed time path formalism for non-equilibrium systems has been developed to study fluctuating hydrodynamics \cite{Grozdanov:2013dba, Kovtun:2014hpa, Harder:2015nxa, Crossley:2015evo, Haehl:2015uoc, Jensen:2017kzi} (see \cite{Glorioso:2018wxw} for a review). The approach allows one to systematically construct the full hydrodynamic effective action including fluctuations about the state of thermal equilibrium, taking into account the effects of non-linear interactions between hydrodynamic modes, purely based on the symmetries of the system. Of particular importance is the dynamical Kubo-Martin-Schwinger (KMS) condition \cite{Crossley:2015evo} used in constructing the effective action, which ensures that the retarded correlation functions of hydrodynamic variables that follow from the action have the correct analyticity properties in the static $\omega \rightarrow 0$ limit \cite{Jain:2020hcu}, which is of interest for the present work.} For instance, the energy-momentum tensor $T^{\mu\nu}$ is sourced by the background metric $g_{\mu\nu}$, and the $U(1)$ current $J^\mu$ is sourced by the corresponding background gauge field $A_\mu$. The hydrodynamic equations, corresponding to energy-momentum and current conservation, 
\begin{equation}
\nabla_\mu T^{\mu\nu} = F^{\nu\lambda} J_\lambda\, ,
\quad \nabla_\mu J^{\mu} = 0\, ,
\end{equation}
follow as a consequence of the diffeomorphism and gauge invariance of the generating functional. Here $F_{\mu\nu} = \partial_\mu A_\nu - \partial_\nu A_\mu$ is the field strength tensor associated with the background field $A_\mu$. The gauge invariance of the generating functional and the ensuing conservation of the current $J^\mu$ can get violated in case the underlying microscopic system has quantum anomalies, which need to be accounted for in the macroscopic hydrodynamic description appropriately \cite{Son:2009tf, Neiman:2010zi}. 

When the background metric and the gauge field vary slowly over length scales much larger than the typical microscopic length scales associated to the system in thermal equilibrium, such as the mean free path or the static correlation length, the generating functional $\mathrm{W}$ admits a derivative expansion in the hydrodynamic variables and the sources. The various coefficients that enter this derivative expansion are called \textit{thermodynamic susceptibilities}, which quantify the response of the equilibrium  system to the presence of external time-independent sources. One can obtain the equilibrium one-point functions for the energy-momentum tensor and the current by varying the static generating functional with respect to the sources:
\begin{equation}
\langle T^{\mu\nu} \rangle = \frac{2}{\sqrt{-g}} \frac{\delta \mathrm{W}}{\delta g_{\mu\nu}}\,\, , \qquad \langle J^\mu \rangle = \frac{1}{\sqrt{-g}} \frac{\delta \mathrm{W}}{\delta A_{\mu}}\, .
\label{eq:1ptdef}
\end{equation}
The resulting constitutive relations for $T^{\mu\nu}$ and $J^\mu$ automatically inherit a derivative expansion, and the coefficients that appear are the transport parameters of the system in equilibrium. Thus, in thermal equilibrium, the susceptibilities appearing in the generating functional are the fundamental objects associated to the system, with the transport parameters entering the constitutive relations being linear combinations of the susceptibilities and their derivatives \cite{Banerjee:2012iz, Jensen:2012jh}.\footnote{An interesting physical consequence of the susceptibilities is to modify matter-gravity equilibrium configurations in general relativity, going beyond the perfect fluid approximation commonly used for such configurations \cite{Kovtun:2019wjz}.} 

For a relativistic fluid in thermal equilibrium, there are no first order terms in the generating functional. However, for a $U(1)$ charged fluid in 3+1 dimensions, there are nine independent thermodynamic susceptibilities at the second order in derivatives \cite{Banerjee:2012iz}. The generating functional has the form \cite{Kovtun:2018dvd}
\begin{equation}
    \mathrm{W}[g_{\mu\nu}, A_\mu] = \int d^4x \sqrt{-g} \left( p(T, \mu) + \sum_{n=1}^9 f_n(T,\mu) \, \mathcal{S}_n\right),
    \label{eq:eqgenfunc}
\end{equation}
where $p(T,\mu)$ is the equilibrium pressure of the fluid, the only zeroth order contribution to the generating functional. The $\mathcal{S}_n$ are nine second order diffeomorphism and gauge invariant quantities constructed from the sources. These are 
\begin{equation}
\label{eq:invariants}
\begin{split}
        \mathcal{S}_1 &= R\, , \quad \mathcal{S}_2 = a^2\, , \quad \mathcal{S}_3 = \Omega^2\, , \quad \mathcal{S}_4 = B^2\, , \quad \mathcal{S}_5 = B\cdot \Omega\, , \\ &\mathcal{S}_6 = E^2\, , \quad \mathcal{S}_7 = E\cdot a\, , \quad \mathcal{S}_8 = B\cdot E\, , \quad \mathcal{S}_9 = B\cdot a.
\end{split}
\end{equation}
Here $R$ is the Ricci scalar for the background metric $g_{\mu\nu}$, $a^\mu \equiv u^\nu \nabla_\nu u^\mu$ is the acceleration and $\Omega^\mu \equiv \epsilon^{\mu\nu\alpha\beta} u_\nu \nabla_\alpha u_\beta$ is the vorticity vector. The electric and magnetic fields are given by $E^\mu \equiv F^{\mu\nu} u_\nu$ and $B^\mu = \frac{1}{2} \epsilon^{\mu\nu\alpha\beta} u_\nu F_{\alpha\beta}$. Note that $a^\mu, \Omega^\mu, E^\mu, B^\mu$ are all orthogonal to the velocity $u^\mu$. The parity, charge conjugation and time reversal properties of the second order invariants are presented in table \ref{table:cpt}.

\begin{table}
\begin{center}
\def\arraystretch{1.2}
\setlength\tabcolsep{4pt}
\begin{tabular}{|c|c|c|c|c|c|c|c|c|c|}
\hline
$\mathcal{S}_n$ & $R$ & $a^2$ & $\Omega^2$ & $B^2$ & $B\cdot\Omega$ & $E^2$ & $E\cdot a$ & $B\cdot E$ & $B\cdot a$\\ \hline\hline
$\mathbb{P}$ & $+$ & $+$ & $+$ & $+$ & $-$ & $+$ & $-$ & $-$ & $+$ \\ \hline
$\mathbb{C}$ & $+$ & $+$ & $+$ & $+$ & $+$ & $+$ & $+$ & $+$ & $+$ \\ \hline
$\mathbb{T}$ & $+$ & $+$ & $+$ & $+$ & $+$ & $+$ & $+$ & $-$ & $-$ \\ \hline
\end{tabular}
\end{center}
\caption{Parity $\mathbb{P}$, charge conjugation $\mathbb{C}$, and time reversal $\mathbb{T}$ eigenvalues for the nine second order equilibrium invariants for a $U(1)$ charged relativistic fluid in 3+1D. Since the $U(1)$ charge we have for the super-Yang-Mills plasma is axial, the electric and magnetic fields that appear here are taken to be axial as well. If the $U(1)$ symmetry is vector in nature, then the seven $\mathbb{T}$-even susceptibilities $f_1, \ldots f_7$ are $\mathbb{P}$-even as well.}
\label{table:cpt}
\end{table}

The $f_n(T,\mu)$ that appear as coefficients of various second order invariants in the equilibrium generating functional eq.\ \eqref{eq:eqgenfunc} are the thermodynamic susceptibilities of the system. These susceptibilities, along with the equilibrium pressure $p$, characterize and capture the equilibrium response of the fluid to external sources, up to the second order in derivatives. Of the nine second order susceptibilities $f_1, \ldots f_9$, seven are time-reversal $\mathbb{T}$ invariant, for both the cases of a vector or an axial $U(1)$ charge. These are the susceptibilities $f_1, \ldots f_7$. Kubo formulas for these seven susceptibilities in terms of the equilibrium two-point functions of the energy-momentum tensor and the $U(1)$ current were found in \cite{Kovtun:2018dvd}. These equilibrium two-point functions are defined via the variation of the one-point functions eq.\ \eqref{eq:1ptdef} with respect to time-independent (or zero-frequency) sources as follows:
\begin{equation}
    \begin{split}
        &\delta_g(\sqrt{-g} \, \langle T^{\mu\nu} \rangle) = \frac{1}{2}\, G_{T^{\mu\nu} T^{\alpha\beta}} (\omega = 0, {\bf{k}})\, \delta g_{\alpha\beta}({\bf{k}})\, ,\\
        &\delta_g(\sqrt{-g} \, \langle J^{\mu} \rangle) = \frac{1}{2}\, G_{J^{\mu} T^{\alpha\beta}} (\omega = 0, {\bf{k}})\, \delta g_{\alpha\beta}({\bf{k}})\, ,\\
        &\delta_A(\sqrt{-g} \, \langle J^{\mu} \rangle) = G_{J^{\mu} J^{\nu}} (\omega = 0, {\bf{k}})\, \delta A_{\nu}({\bf{k}})\, .
    \end{split}
\end{equation}
These lead to the Kubo formulas presented in table \ref{tableKubo} for the $\mathbb{T}$-even second order susceptibilities $f_1, \ldots f_7$. The $\mathbb{T}$-odd susceptibilities $f_8, f_9$ unfortunately do not admit Kubo formulas in terms of equilibrium two-point functions, and will not be of interest in the subsequent discussion. 
\begin{table}[h!]
\centering
 \begin{tabular}{|c||c|}
 \hline
 \textbf{Susceptibility} & \textbf{Kubo formula} \\
 \hline\hline
 $f_1$ & $- \frac{1}{2} \lim\limits_{{\bf k} \rightarrow 0} \frac{\partial^2}{\partial k_z^2} G_{T^{xy} T^{xy}}$ \\[5pt]
 \hline
 $f_2$ & $\frac{1}{4} \lim\limits_{{\bf k} \rightarrow 0} \frac{\partial^2}{\partial k_z^2} \left( G_{T^{tt} T^{tt}} + 2 G_{T^{tt} T^{xx}} -4 G_{T^{xy} T^{xy}}\right)$ \\[5pt]
 \hline
 $f_3$ &  $\frac{1}{4} \lim\limits_{{\bf k} \rightarrow 0} \frac{\partial^2}{\partial k_z^2} \left( G_{T^{tx} T^{tx}} + G_{T^{xy} T^{xy}} \right)$ \\[5pt]
 \hline
 $f_4$ & $\frac{1}{4} \lim\limits_{{\bf k} \rightarrow 0} \frac{\partial^2}{\partial k_z^2} G_{J^x J^x}$  \\[5pt]
 \hline
 $f_5$ & $\frac{1}{2} \lim\limits_{{\bf k} \rightarrow 0} \frac{\partial^2}{\partial k_z^2} G_{J^x T^{tx}}$  \\[5pt]
 \hline
 $f_6$ &  $\frac{1}{4} \lim\limits_{{\bf k} \rightarrow 0} \frac{\partial^2}{\partial k_z^2} G_{J^t J^t}$ \\[5pt]
 \hline
 $f_7$ &   $- \frac{1}{2} \lim\limits_{{\bf k} \rightarrow 0} \frac{\partial^2}{\partial k_z^2} \left( G_{J^{t} T^{tt}} + G_{J^{t} T^{xx}} \right)$ \\[5pt]
 \hline
 \end{tabular}
 \caption{List of Kubo formulas for $\mathbb{T}$-even second order thermodynamic susceptibilities $f_1, \ldots f_7$.}
 \label{tableKubo}
\end{table}

As mentioned above, the holographic description of the charged strongly coupled $\mathcal{N} = 4$ super-Yang-Mills plasma in thermal equilibrium in the absence of background electric and magnetic fields is provided by the Reissner-Nordstr\"{o}m black brane solution in asymptotically AdS$_5$ spacetime \cite{Chamblin:1999tk, Cvetic:1999ne}. By perturbing this geometry and the gauge field in the bulk, one can compute the retarded two-point functions of the energy-momentum tensor and the $U(1)$ current for the boundary theory purely from the bulk perspective. By computing these retarded correlation functions and making use of the Kubo formulas, it is straight forward to extract the seven $\mathbb{T}$-invariant thermodynamic susceptibilities $f_1,\ldots f_7$ for the strongly coupled $\mathcal{N} = 4$ super-Yang-Mills plasma. In fact, the behaviour of these second order susceptibilities as functions of the parameters $\mu$ and $T$, extracted using holography, is one of the main results of this paper. These results for the equilibrium transport properties of the super-Yang-Mills plasma serve as a proxy for the behaviour of more realistic strongly interacting systems found in nature, such as strongly coupled QCD matter at a non-zero baryon number chemical potential.\footnote{As applications of the Kubo formulas mentioned, non-vanishing second-order susceptibilities for some free quantum field theories at finite temperature but zero chemical potential were  computed in \cite{Kovtun:2018dvd}. Also, \cite{Shukla:2019shf} provides analytic results for the susceptibilities for massive Dirac fermions which are degenerate \textit{i.e.}\ at a finite chemical potential but vanishing temperature, as well as numerical results for when both temperature and chemical potential are non-zero.}

Let us briefly comment upon some of the known aspects of second order transport for the strongly coupled $\mathcal{N} = 4$ super-Yang-Mills plasma, which is a conformal theory. For the uncharged state of the plasma, there are five transport parameters at the second order in the derivative expansion. These are usually denoted by $\tau_\pi, \kappa, \lambda_1, \lambda_2$ and $\lambda_3$, and in the limit of infinite 't-Hooft coupling are given by \cite{Baier:2007ix, Bhattacharyya:2008jc}
\begin{equation}
    \tau_\pi = \frac{2-\ln 2}{2\pi T}\, , \quad \kappa = \frac{\eta}{\pi T}\, , \quad \lambda_1 = \frac{\eta}{2\pi T}\, , \quad \lambda_2 = - \frac{\eta \ln 2}{\pi T}\, , \quad \lambda_3 = 0.
    \label{n4neutral}
\end{equation}
Here $\eta$ is the shear viscosity of the plasma, which is the only transport parameter at the first order for the uncharged state, and is given by $\eta = \frac{\pi}{8} N_c^2 T^3$ \cite{Policastro:2001yc}.\footnote{References studying first order transport for the charged $\mathcal{N} = 4$ super-Yang-Mills plasma, where one also has the conductivity along with the shear viscosity as a transport parameter, at a finite chemical potential, include \cite{Mas:2006dy, Son:2006em, Maeda:2006by, Benincasa:2006fu, Ge:2008ak, Matsuo:2009yu, Myers:2009ij, Cremonini:2009sy, Matsuo:2009xn, Kontoudi:2012mu}. See also \cite{Sahoo:2009yq} for a discussion of the dispersion relations of the super-Yang-Mills plasma, and how they get affected by the presence of the anomaly.} Note that $\kappa$ and $\lambda_3$ are thermodynamic in nature, and are related to our $f_1, f_3$ via $\kappa = -2 f_1$ and $\lambda_3 = 2\left(T \frac{\partial f_1}{\partial T} + \mu \frac{\partial f_1}{\partial \mu} - 4 f_3\right)$ \cite{Kovtun:2018dvd}. Interestingly, the parameters $\lambda_1, \lambda_2$ and $\tau_\pi$ satisfy the Haack-Yarom relation \cite{Haack:2008xx}
\begin{equation*}
4\lambda_1 + \lambda_2 = 2\eta \tau_\pi,   
\end{equation*}
which is satisfied by a large class of conformal gauge theory plasmas. Leading corrections to the results in eq.\ \eqref{n4neutral} ensuing from a finite but large value of the `t-Hooft coupling $\lambda$ have been found in \cite{Buchel:2008bz, Buchel:2008kd, Saremi:2011nh, Grozdanov:2014kva}. In particular, \cite{Grozdanov:2014kva} argues that the Haack-Yarom relation holds true even when leading coupling constant corrections are taken into account. Finally, for the $U(1)$ charged super-Yang-Mills plasma, there are eight additional transport parameters at the second order. This entire set of thirteen second order parameters was computed in \cite{Erdmenger:2008rm, Banerjee:2008th} using the fluid/gravity correspondence \cite{Bhattacharyya:2008jc}. As mentioned earlier, in thermal equilibrium, the generating functional and the susceptibilities provide the fundamental description of the macroscopic state of the system, with the transport parameters being derived quantities. We therefore obtain the second order susceptibilities directly in the present article. Other important references for transport in $\mathcal{N} = 4$ super-Yang-Mills plasma include \cite{Heller:2007qt, Arnold:2011ja}. 

The paper is organized as follows. In section \ref{holo_model}, we provide details of the holographic geometry dual to the charged strongly coupled $\mathcal{N} = 4$ super-Yang-Mills plasma in thermal equilibrium, \textit{i.e.} the Reissner-Nordstr\"{o}m black brane solution in asymptotically AdS$_5$ spacetime. In section \ref{computations} we provide details of the procedure used to extract the seven $\mathbb{T}$-even second order susceptibilities $f_1, \ldots f_7$ for the charged super-Yang-Mills plasma. We also compare our results to some of the things previously known in literature, and also make an estimate of the transport parameter $\kappa$ for QCD using our holographic results, comparing it with previous determinations made using different techniques. We end with some concluding comments in section \ref{discussion}. Appendices provide supplementary material relevant for the discussion in the main body of the paper, including details of the numerical procedure in appendix \ref{app:numericalmethods}. 

\textit{\underline{Notation and conventions}:} The metric signature is mostly positive. Capitalized Latin letters denote 4+1 dimensional bulk spacetime indices: $A, B,\dots = 0,1,2,3,4$. Greek letters denote 3+1 dimensional boundary spacetime indices: $\mu, \nu, \ldots = 0,1,2,3$. Latin small letters denote boundary spatial directions: $i, j, \ldots = 1,2,3$. Spatial three-vectors are denoted by bold letters e.g.\ ${\bf k, x}$ etc.

\section{The holographic model}
\label{holo_model}
As per the AdS/CFT correspondence \cite{Maldacena:1997re, Gubser:1998bc, Witten:1998qj}, the four-dimensional $\mathcal{N} = 4$ supersymmetric $SU(N_c)$ Yang-Mills theory is holographically dual to type IIB superstring theory on $\text{AdS}_5 \times S^5$. Further, in the limit where $N_c \rightarrow \infty$ and the gauge theory is strongly coupled $g_{\textrm{YM}}^2 N_c \rightarrow \infty$, the dual bulk description simplifies considerably and is provided by classical type IIB supergravity. By appropriately compactifying the ten-dimensional supergravity theory on $S^5$, one obtains a consistent truncation to the five-dimensional Einstein-Maxwell-Chern-Simons theory \cite{Chamblin:1999tk, DHoker:2009mmn, DHoker:2009ixq}, given by the action\footnote{The consistent truncation of IIB supergravity on $\text{AdS}_5\times S^5$ gives the bulk gauge field coupling to be $g^2 = \frac{2\kappa_5^2}{L^2}$. See for instance \cite{Chamblin:1999tk}. We have absorbed a factor of $L$ in defining the bulk gauge field to make it dimensionless.}
\begin{equation}
\label{eq:bulk_action}
    S = \int_{\mathcal{M}}\dd^5x\,\sqrt{-\mathcal{G}}\left(\frac{1}{2\kappa_5^2}\left(\mathcal{R}-2\Lambda\right)-\frac {1}{4g^2}\, \mathcal{F}_{MN} \mathcal{F}^{MN}\right)- \frac{\alpha}{6 g^2} \int\dd^5x\,\left(\mathcal{A} \wedge \mathcal{F} \wedge \mathcal{F}\right) + S_\text{bdy},
\end{equation}
where $\mathcal{R}$ is the Ricci scalar corresponding to the five-dimensional bulk metric $\mathcal{G}_{MN}$, and $\mathcal{F}_{MN}=\partial_M \mathcal{A}_N-\partial_N \mathcal{A}_M$ is the field strength tensor for the bulk Abelian gauge field $\mathcal{A}_M$.  Also, $\kappa_5^2 = 8\pi G_N$, and the cosmological constant is related to the AdS radius $L$ via $\Lambda = - 6/L^2$. The coefficient of the Chern-Simons term, $\alpha$, has the numerical value $2/\sqrt{3}$, and is a measure of the strength of the chiral anomaly in the boundary super-Yang-Mills theory.\footnote{The presence of the Chern-Simons term is also required by supersymmetry in $\mathcal{N} = 2$ gauged supergravity theory on $\textrm{AdS}_5$, which has precisely the same action for its bosonic sector as the one in eq.\ \eqref{eq:bulk_action} \cite{Freedman:1976aw, Romans:1991nq, Cadavid:1995bk, Papadopoulos:1995da, Buchel:2006gb}.}

The action eq.\ \eqref{eq:bulk_action} is supplemented by boundary- and counter-terms \cite{Henningson:1998gx,  Balasubramanian:1999re, Emparan:1999pm, Taylor:2000xw, Batrachenko:2004fd, Sahoo:2010sp},
\begin{equation}
S_{\text{bdy}}= \frac{1}{\kappa_5^2} \int\limits_{\partial\mathcal{M}}\! \dd^4x  \sqrt{-\gamma} \left( K - \frac{3}{L} - \frac{L}{4} R +  \log\left( \frac{u}{L} \right) \left( \frac{\kappa_5^2\,L}{4\,g^2} F_{\mu\nu} F^{\mu\nu} - \frac{L^3}{8} R_{ \mu\nu} R^{\mu\nu} +  \frac{L^3}{24} {{R}}^{2} \right)\right) ,
\end{equation}
where $u$ denotes the radial direction in Poincar\'e slicing.
The metric $\gamma_{\mu\nu}$ is induced by the bulk metric $\mathcal{G}_{MN}$ on the conformal boundary of AdS$_5$, and $R$ is the associated four-dimensional boundary Ricci scalar. Furthermore, $K_{MN}$ is the extrinsic curvature, given by
\begin{equation}
\label{eq:extrinsicK}
K_{MN} = \mathcal{P}_M^{\ \, L} \,  \mathcal{P}_N^{\ \, P} \, \nabla_L n_P \, ,\qquad \mbox{with} \quad \mathcal{P}_M^{\ \, L} = \delta_M^{\ L} - n_M n^L \, ,
\end{equation} 
with $\nabla$ being the covariant derivative and $n^M$ being the outward pointing normal vector to the boundary $\partial\mathcal{M}$.

The equations of motion for the metric $\mathcal{G}_{MN}$ and the gauge field $\mathcal{A}_M$ read
\begin{align}
   & \mathcal{R}_{MN}-\frac{\mathcal{R}}{2} \mathcal{G}_{MN}-\frac {6}{L^2} \mathcal{G}_{MN}=\frac{\kappa_5^2}{g^2}\left(\mathcal{F}_{MJ} \mathcal{F}^J_N - \frac 14 \mathcal{G}_{MN}\,\mathcal{F}_{JK} \mathcal{F}^{JK}\right), \label{eq:eom1}\\
   &\nabla_M \mathcal{F}^{MN} +\frac{\alpha}{8\,\sqrt{-\mathcal{G}}}\,\epsilon^{NMOPQ}\mathcal{F}_{MO}\mathcal{F}_{PQ} = 0. \label{eq:eom2}
\end{align}
Here $\epsilon^{NMOPQ}$ is the totally anti-symmetric Levi-Civita symbol in $4+1$ dimensions, with $\epsilon^{txyzu}=1$.

The four-dimensional $\mathcal{N} = 4$ supersymmetric $SU(N_c)$ Yang-Mills theory carries a global $SU(4)$ R-symmetry. From the AdS/CFT perspective, this is same as the $SO(6)$ symmetry of the dual type IIB supergravity theory on $\textrm{AdS}_5 \times S^5$, corresponding to the isometries of $S^5$. The effective theory eq.\ \eqref{eq:bulk_action} can be obtained from supergravity by simply providing rotations or twists to the angular directions on $S^5$ and compactifying, as illustrated beautifully in \cite{Chamblin:1999tk}. In fact, their construction corresponds to introducing a rotation in the diagonal $U(1)$ of the maximal Abelian subgroup $U(1)^3$ of the R-symmetry group. For the super-Yang-Mills theory, this corresponds to considering states with a non-vanishing expectation value for the dual $U(1)$ current $J^\mu$. This precisely is the charged super-Yang-Mills plasma whose transport properties are of our interest here. In particular, the equilibrium state of the plasma at a finite temperature and chemical potential, in the absence of background electric and magnetic fields, is holographically dual to the Reissner-Nordstr\"{o}m black brane solution for the action eq.\ \eqref{eq:bulk_action} \cite{Chamblin:1999tk, Cvetic:1999ne}, which in infalling Eddington-Finkelstein coordinates takes the form \cite{ammon_erdmenger_2015, Erdmenger:2008rm, Banerjee:2008th}
\begin{equation}
    \dd s^2=\frac{L^2}{u^2}\,\left(-f(u)\,\dd t^2-2\,\dd t\,\dd u+\dd x^2+\dd y^2+\dd z^2\right).\label{eq:metric}
\end{equation}
Here, for brevity, we denote the null ingoing Eddington-Finkelstein time coordinate by $t$, while the radial coordinate is denoted by $u$. The horizon of the black brane is located at $u=u_h$, and the temperature of the dual field theory is given by the black brane temperature, i.e.\ $T = |f'(u_h)|/4\pi$. The boundary is located at $u=0$, where we impose asymptotically AdS$_5$ boundary conditions, i.e.\ $f(0)=1$.
Also, the bulk gauge field $\mathcal{A}_M$ takes the form
\begin{equation}
    \mathcal{A} = \mathcal{A}_t(u)\,\dd t,
\end{equation}
with other components vanishing. We fix our gauge by requiring that $\mathcal{A}_t(u_h)=0$ and by imposing $\mathcal{A}_u(u)=0$. In this gauge, the chemical potential of the boundary theory is simply the boundary value of $\mathcal{A}_t$ i.e.\ $\mu = \mathcal{A}_t(0)$. 
The equations of motion, eqs.\ \eqref{eq:eom1} and \eqref{eq:eom2}, are solved by the Reissner-Nordstr\"om AdS$_5$ black brane metric \cite{Erdmenger:2008rm, Banerjee:2008th, ammon_erdmenger_2015}, 
\begin{equation}
    f(u)=1-\left(1+\frac{2\,u_h^2\,\mu^2\,\kappa_5^2}{3\,L^2\,g^2}\right)\left(\frac{u}{u_h}\right)^4+\frac{2\,u_h^2\,\mu^2\,\kappa_5^2}{3\,L^2\,g^2}\left(\frac{u}{u_h}\right)^6,\label{eq:blackening}
\end{equation}
with the gauge field given by
\begin{equation}
 \mathcal{A}_t(u) = \mu\left(1-\frac{u^2}{u_h^2}\right).
\end{equation}

From eqs.\ \eqref{eq:metric} and \eqref{eq:blackening}, it is straightforward to compute the temperature $T$ and entropy density $s$, which are given by
\begin{align}
    &T=\frac{|f'(u_h)|}{4\pi}=\frac{1}{\pi\,u_h}\left(1-\frac{u_h^2\,\mu^2\,\kappa_5^2}{3\,L^2\,g^2}\right),\label{eq:temp}\\
    &s=\frac{A_\text{BH}}{4\,G_N\,\text{Vol}(\mathbb R^3)}=\frac{2\pi}{\kappa_5^2}\left(\frac{L}{u_h}\right)^3,
\end{align}
where $A_\text{BH}$ is the area of the black brane horizon.

As discussed in section \ref{Introduction}, the thermodynamic susceptibilities $f_1, \ldots f_7$ are given by Kubo formulas involving equilibrium two-point functions of the energy-momentum tensor and the $U(1)$ current (see table \ref{tableKubo}). Now the energy-momentum tensor of the boundary theory is sourced by the bulk metric, while the $U(1)$ current is sourced by the bulk gauge field. One can thus extract the two-point functions for the energy-momentum tensor and the $U(1)$ current by studying perturbations to the background  metric and gauge field configurations discussed above. For our purposes, it is useful to consider time-independent spatial fluctuations on top of the background solution $(\mathcal{G}_{MN}, \mathcal{A}_M)$, \textit{i.e.}
\begin{equation}
\label{eq:perturbed}
    \tilde{\mathcal{G}}_{MN}=\mathcal{G}_{MN}(u)+\epsilon\, h_{MN}(u)\,e^{i\,k_z\, z}+\mathcal O(\epsilon^2)\, ,\quad\quad \tilde{\mathcal{A}}_M(u)= \mathcal{A}_M(u)+\epsilon\, a_M(u) \,e^{i\,k_z\, z}+\mathcal O(\epsilon^2)\, ,
\end{equation}
to first order in a small parameter $\epsilon$. To fix our gauge at the level of fluctuations, we work in the radial gauge by setting $h_{Mu},a_u$ to zero.
Due to the isotropy of the system, without loss of generality  we have chosen the momentum to point along the $z$-direction, and will drop the subscript $z$ for the remainder of this work \textit{i.e.}\ $k \equiv k_z$.

\section{Computing the thermodynamic susceptibilities}
\label{computations}
In this section, we compute the seven $\mathbb{T}$-even second order thermodynamic susceptibilities $f_1, \ldots f_7$  for the $\mathcal{N} = 4$ super-Yang-Mills plasma making use of the holographic perturbed Reissner-Nordstr\"{o}m black brane model described in the previous section.

We may compute holographic two-point functions of the form $G_{\mathcal O^a\mathcal O^c}(k)$ in terms of the expectation value of the one-point function $\langle \mathcal{O}^a\rangle$ by exploiting the following relation to the first order in the fluctuations~\cite{Horowitz:2008bn,ammon_erdmenger_2015},
\begin{equation}
    \delta\!\left\langle \mathcal{O}^a \right\rangle(k) = G_{\mathcal{O}^a \mathcal{O}^c}(k) \, \delta \phi_c( k) \, ,\label{eq:onepointtwopoint}
\end{equation}
where the perturbation $\delta \phi_c$ is the source dual to the operator $\mathcal{O}^c$ on the boundary and $\delta\!\left\langle\mathcal{O}^a\right\rangle$ is the response for turning on this perturbation. For the two-point functions we would like to compute, $\mathcal{O}^a \equiv T^{\mu\nu}, J^\mu$, and the sources correspond to perturbations in the bulk metric and the gauge field.
The expressions for the renormalized one-point functions for $T^{\mu\nu}, J^\mu$ follow from the standard holographic renormalization procedure.
The renormalized energy-momentum tensor may be extracted from the covariant expression \cite{Balasubramanian:1999re}
\begin{align}
2\kappa_5^2\,\left\langle T_{\mu\nu} \right\rangle=\lim\limits_{u\rightarrow 0}\frac{1}{u^{2}}&\left[  -2K_{\mu\nu}+2(K-3) \, \gamma_{\mu\nu}+\frac{2\kappa_5^2}{g^2} \log\left(\frac{u}{L}\right)\left(F_{\mu}^{\ \alpha}F_{\nu\alpha}
-\frac{1}{4} \, \gamma_{\mu\nu}F^{\alpha\beta}F_{\alpha\beta}\right)\right.\nonumber\\
&\left.\quad+ R_{\mu\nu} - \frac{1}{2} R \gamma_{\mu\nu} + 8 \log\left(\frac uL\right)  h^{(4)}_{\mu\nu}\right] \, , \label{eq:Tmunu}
\end{align}
with
\begin{equation*}
h^{(4)}_{\mu\nu} =    \frac{1}{8} R_{\mu\nu\rho\sigma} R^{\rho\sigma} + R\, \nabla_\mu \nabla_\nu  R - \frac{1}{16} \nabla^2 R_{\mu\nu} - \frac{1}{24} R R_{\mu\nu} + \frac{1}{96} \left( \nabla^2 R + R^2 - 3 R_{\rho\sigma} R^{\rho\sigma}  \right) \gamma_{\mu\nu}.
\end{equation*}

Similarly, the one-point function of the renormalized current is given by \cite{DHoker:2009ixq, Megias:2013joa, Ammon:2020rvg} \footnote{This is the expression for the covariant current, as opposed to the gauge-dependent consistent current. The two differ by the addition of a Bardeen-Zumino term $ \frac{\alpha}{6} \epsilon^{\mu\nu\rho\sigma} A_\nu F_{\rho\sigma}$. See \cite{Ammon:2020rvg} for a detailed discussion.}
\begin{equation}\label{eq:currentJcov}
g^2\left\langle J^{\mu} \right\rangle= \lim\limits_{u\rightarrow 0}\left(
\sqrt{-\gamma}\, n_M\, \mathcal{G}^{M\nu}F_{\nu\sigma} \gamma^{\sigma \mu} + \log\left(\frac{u}{L}\right)\, \sqrt{-\gamma} \, {\nabla}_\nu F^{\nu\mu} \right) .
\end{equation}

Up to contact terms and terms from the holographic renormalization procedure, the two-point function reads
\begin{equation}
    G_{\mathcal O^a\mathcal O^c}=L^3\,(2\Delta-4)\frac{\langle\mathcal O^a(k)\rangle}{\delta\phi_c(k)},
\end{equation}
where $\Delta$ is the conformal dimension of $\mathcal O^a$.
We obtain the value of the two-point function by computing the expectation value of $\mathcal{O}^a(k)$ in the presence of a source term $\delta\phi_c(k)$ for the operator $\mathcal O^c$ and computing their ratio. 
Note that the asymptotic boundary expansion of the gauge field fluctuations read formally as $a_\mu(u;k)\sim a^{\textbf{L}}_\mu+u^2/L^2\,\left(a^{\textbf{S}}_\mu(u;k)+a^{\textbf{L}}_\mu\,k^2\,\log(u/L)\right)$, where the superscripts \textbf{L} and \textbf{S} denote the leading and subleading modes. According to the holographic dictionary, the expectation value of the dual operator is encoded in the subleading mode of the asymptotic boundary expansion. However, at the order $u^2$, we observe logarithmic divergences which are removed by boundary counter-terms. These inherently break conformal invariance and we have to specify the renormalization scale when regularizing the action (see \cite{Horowitz:2008bn,Fuini:2015hba} for more details). In our context of the second order thermodynamic transport coefficients, this leads to additional finite contributions to the renormalized expectation value at order $k^2$ and affects the current-current two-point functions $G_{J^tJ^t}$ and $G_{J^xJ^x}$. The two-point functions involving the energy-momentum tensor are unaffected at order $k^2$, since for them the logarithmic terms due to the Weyl anomaly are of order $k^4$.

We compute the susceptibility $f_1$ analytically in appendix~\ref{app:analytical} - see eqs.\ \eqref{eq:analyticalresult} and \eqref{eq:f1anal2}. Working in units where we set $g^2=2\kappa_5^2=L=1$ for simplicity, we get the analytic result
\begin{equation}
    f_1 = -\frac{1}{2} \frac{\mu^4}{\left(3\pi T - \sqrt{9\pi^2 T^2 + 6\mu^2}\right)^2}.
    \label{eq:f1anal}
\end{equation}
Now, for a conformal theory, such as $\mathcal{N} = 4$ super-Yang-Mills, the susceptibilities $f_2$ and $f_7$ are not independent of $f_1$ \cite{Kovtun:2018dvd}. In fact, $f_2 = 6f_1$ and $f_7 = -6 \partial f_1/\partial \mu$. These relations between the susceptibilities are required to maintain the Weyl invariance of the equilibrium generating functional $\text{W}[g_{\mu\nu}, A_\mu]$ for the conformal theory. Using eq.\ \eqref{eq:f1anal}, we thus get the following results for $f_2$ and $f_7$,
\begin{align}
    f_2 &= - \frac{3 \mu^4}{\left(3\pi T - \sqrt{9\pi^2 T^2 + 6\mu^2}\right)^2}, \\
    f_7 &= \mu + \frac{\sqrt{3} \pi T \mu}{\sqrt{3\pi^2 T^2 + 2 \mu^2}}. 
\end{align}

\begin{figure}[h!]
    \centering\begin{minipage}{\dimexpr\paperwidth-4cm}
    \includegraphics[width=0.32 \linewidth]{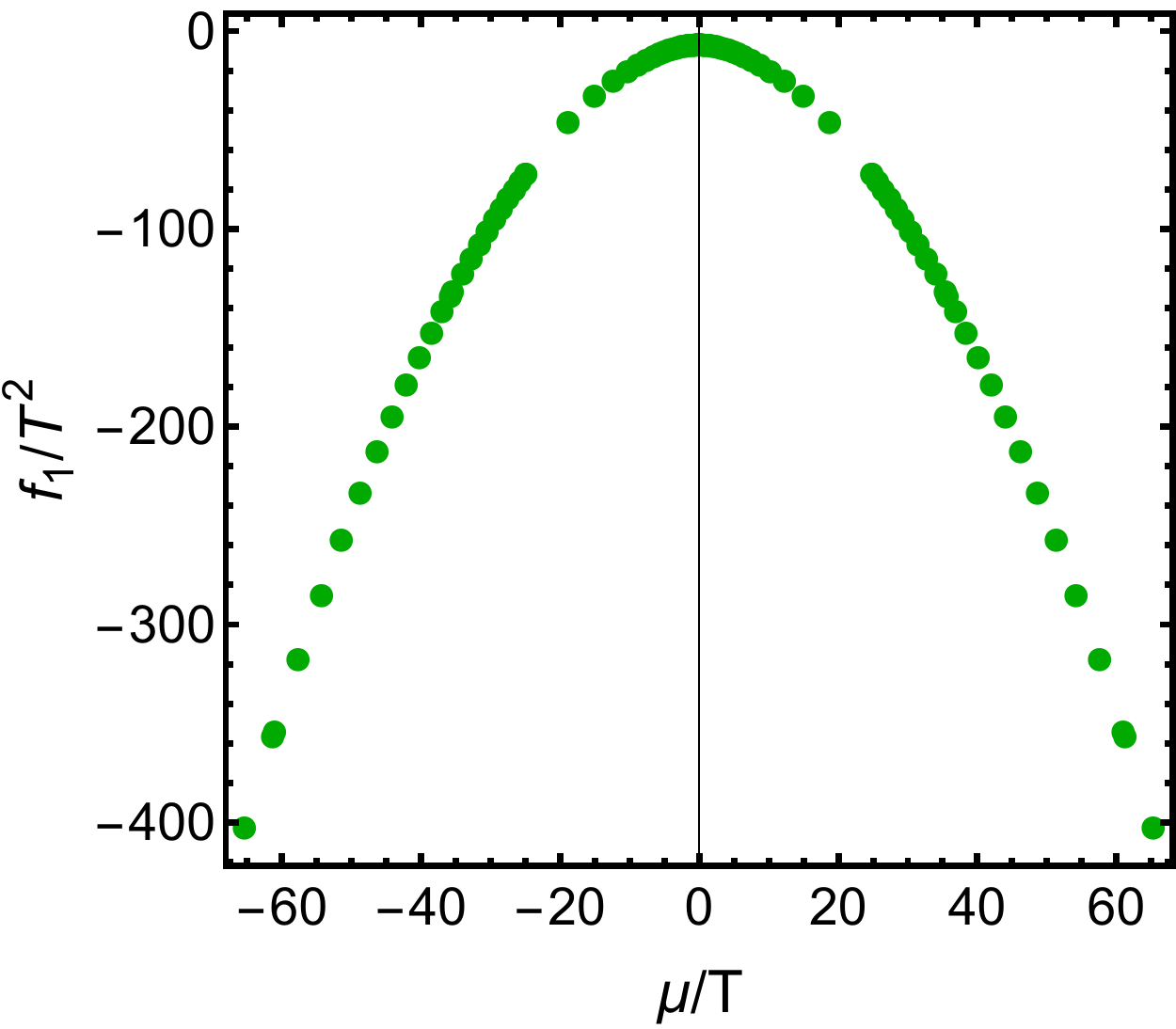} \quad\quad\quad
    \includegraphics[width=0.32 \linewidth]{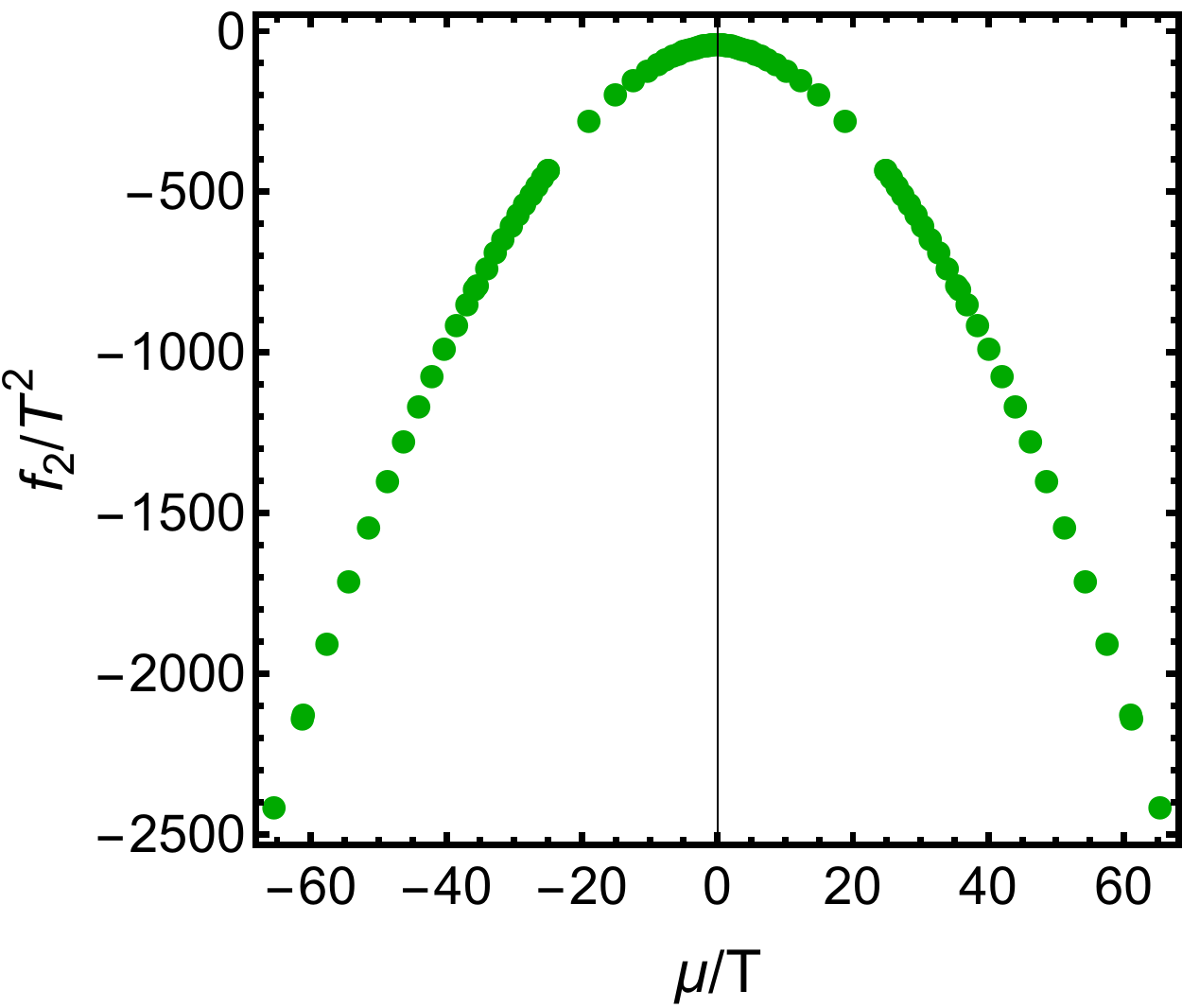}
    \mbox{}\vspace{3mm}
    \includegraphics[width=0.33 \linewidth]{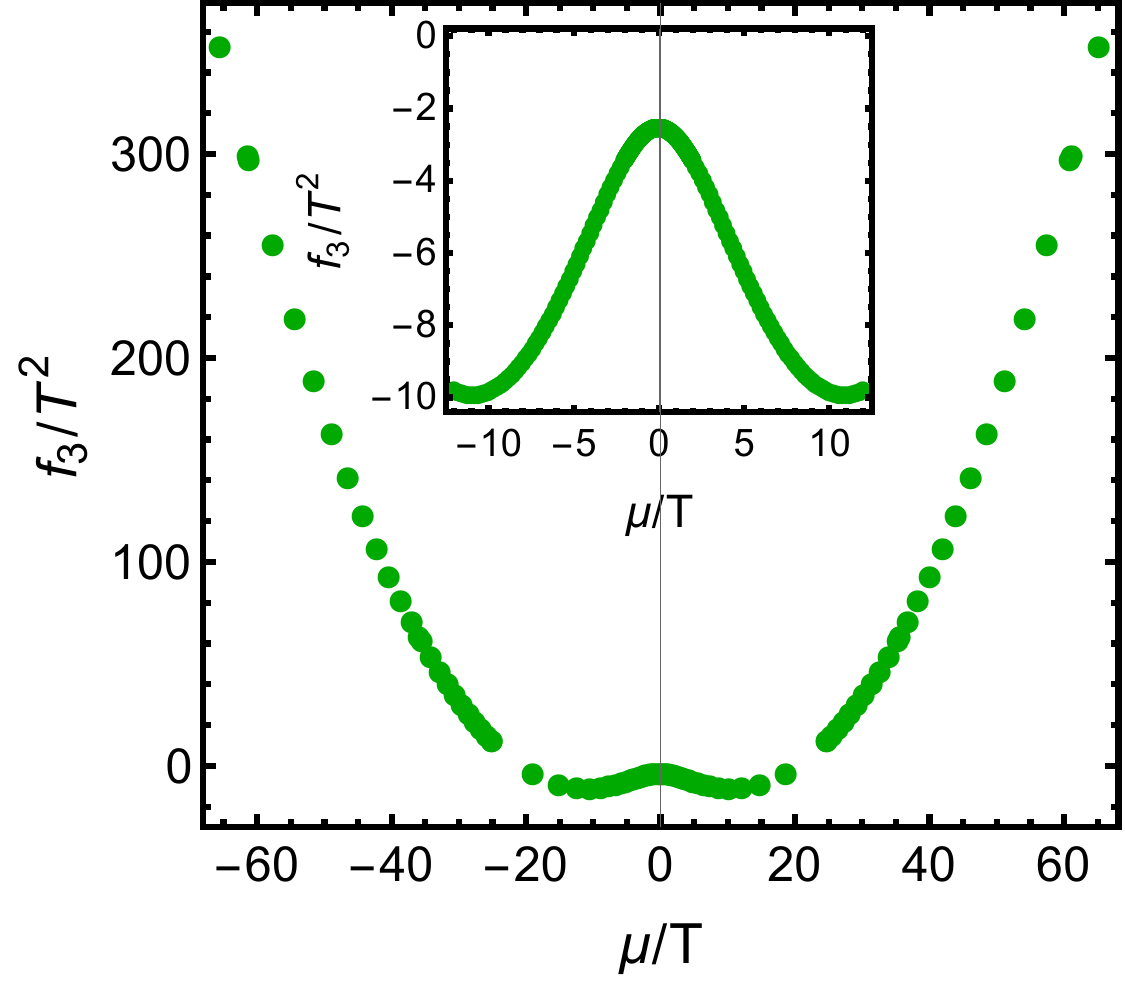} \quad\quad\quad
    \includegraphics[width=0.32 \linewidth]{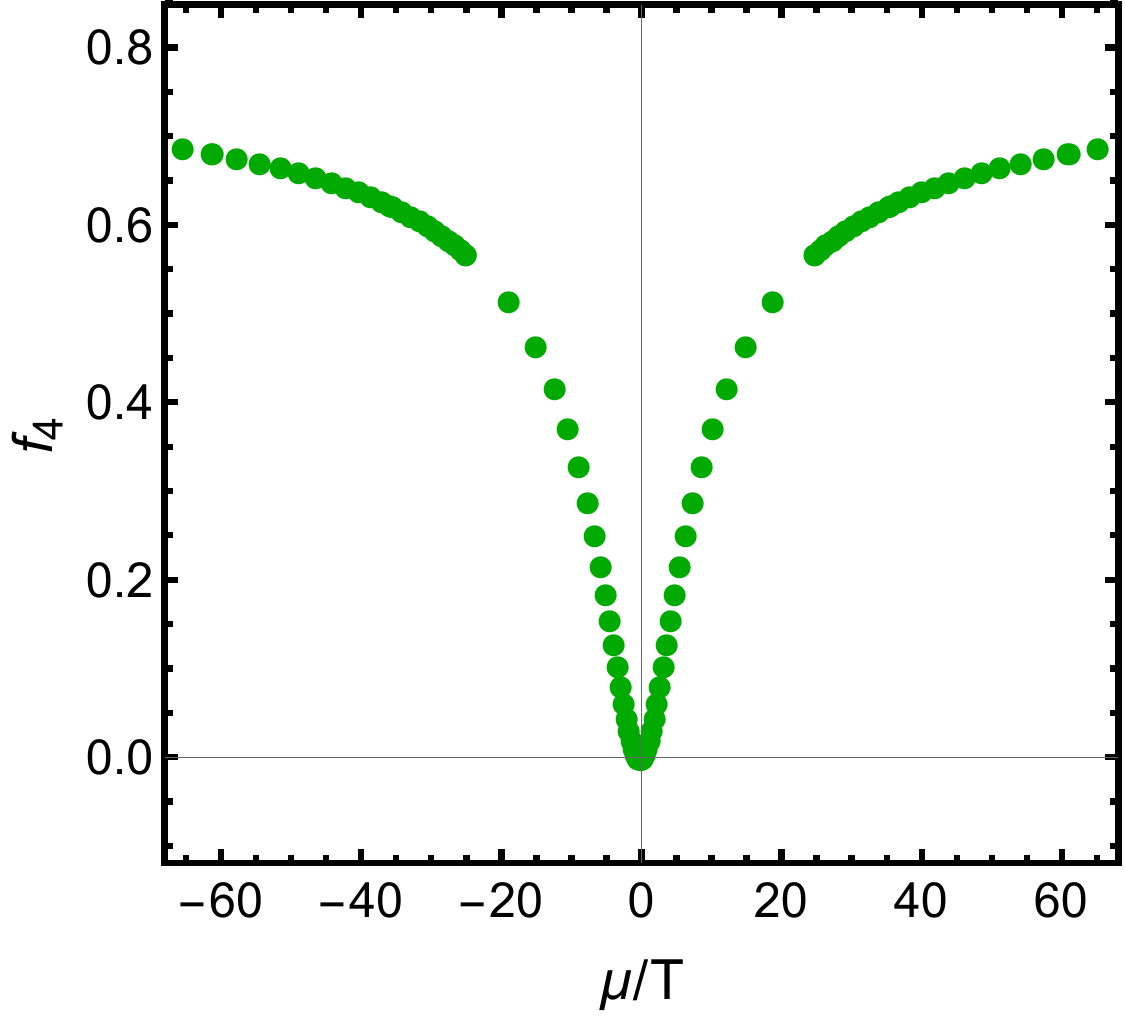}
    \mbox{}\vspace{3mm}
    \includegraphics[width=0.32 \linewidth]{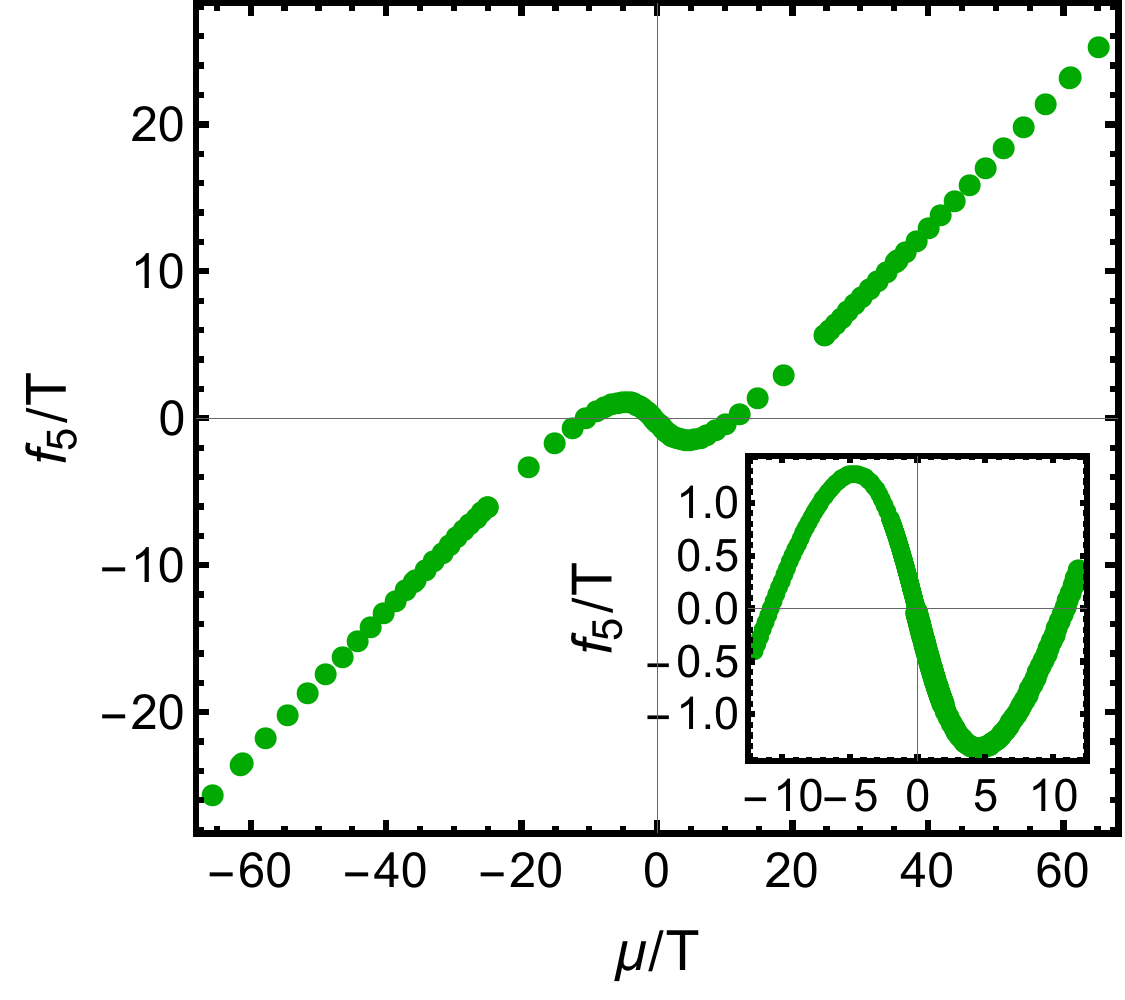} \quad\quad\quad
    \includegraphics[width=0.32 \linewidth]{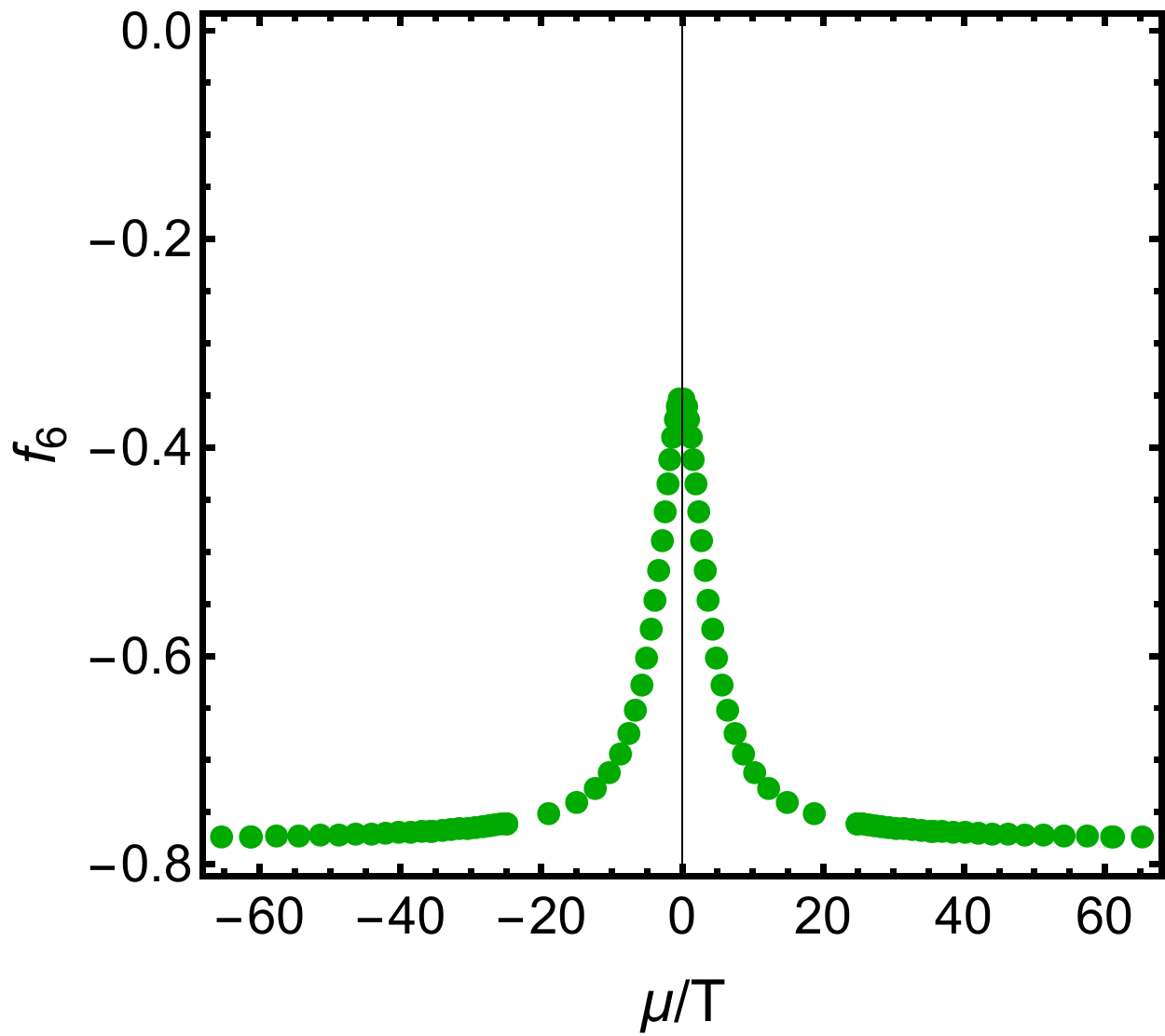}
    \mbox{}\vspace{3mm}
    \includegraphics[width=0.32 \linewidth]{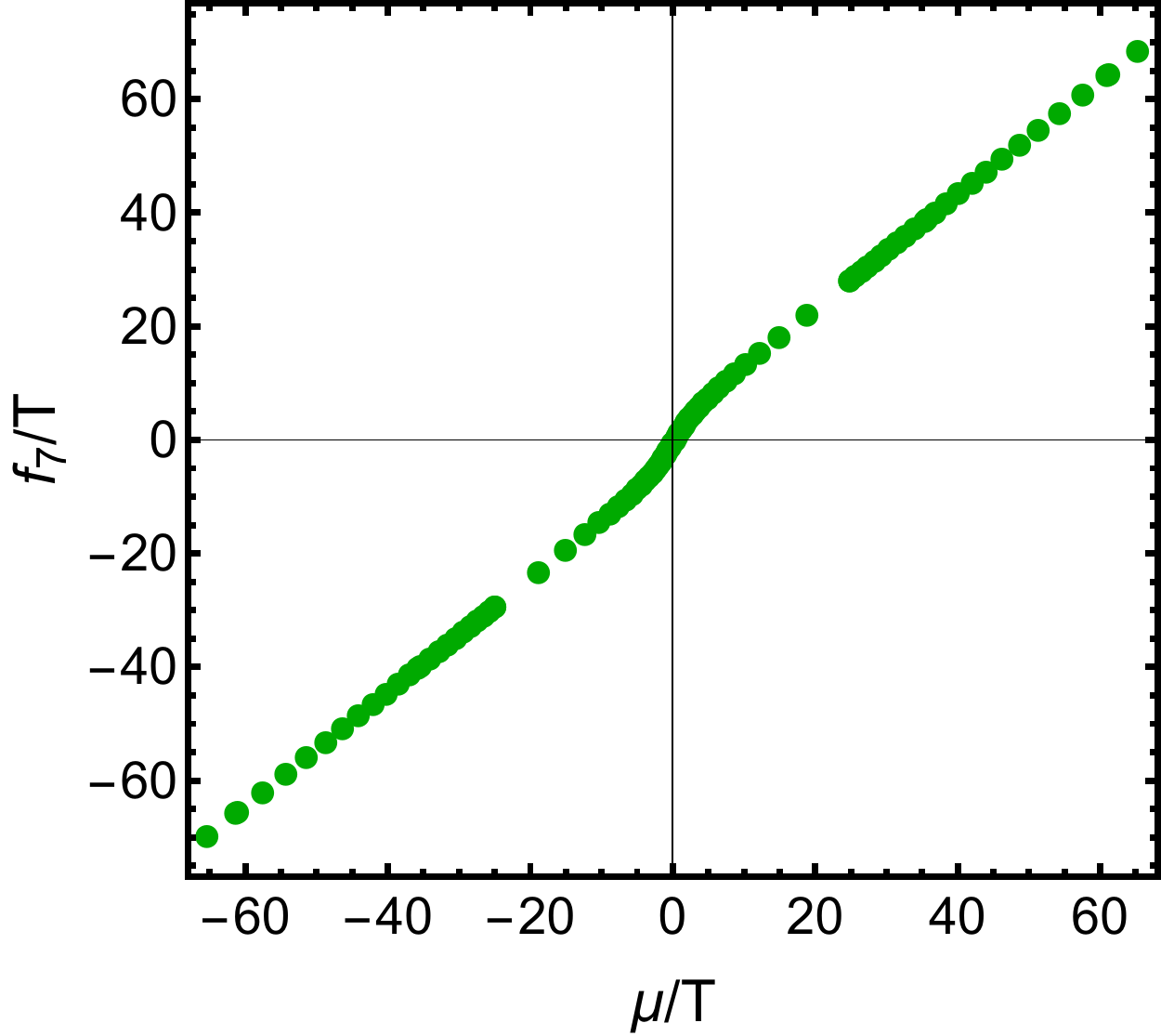}
    \caption{Susceptibilities $f_1, \ldots f_7$ for strongly coupled $\mathcal{N} = 4$ SYM as functions of the dimensionless parameter $\mu/T$. The insets for $f_3, f_5$ depict zoomed-in behaviour in the vicinity of the origin.}
    \label{fig:plots}
    \end{minipage}
\end{figure}

The rest of the susceptibilities are computed numerically by means of pseudo-spectral methods as explained in appendix~\ref{app:numericalmethods}. To compute the second order in $k$ Green's function, we compute the Green's function for several small values of the momentum $k$ and extract the $k^2$ dependence by fitting the numerical results to a second order polynomial (in $k$). Our final results for the seven $\mathbb{T}$-even thermodynamic susceptibilities for the charged strongly-coupled $\mathcal{N} = 4$ supersymmetric $SU(N_c)$ Yang-Mills plasma in the limit of large $N_c$, extracted via the holographic procedure described above, are depicted in figure \ref{fig:plots}. 

Finally, we discuss the behaviour of the susceptibilities under a parity transformation, as well as for small chemical potentials (or high temperatures) $|\mu|/T \ll 1$, summarized in table~\ref{tab:scaling}. The behaviour under a parity transformation matches exactly the $\mathbb{P}$ eigenvalues of the equilibrium invariants given in eq.\ \eqref{eq:invariants}; see table \ref{table:cpt}. For small values of the dimensionless chemical potential $|\mu|/T$, in accordance with the transformation properties under parity, the dimensionless transport coefficients $f_1/T^2,\,f_2/T^2,\, f_3/T^2, f_4$ and $f_6$ are quadratic in $\mu/T$, whereas $f_5/T$ and $f_7/T$ are linear in $\mu/T$. The respective prefactors for their behaviour in the limit $|\mu|/T \ll 1$ are presented in table~\ref{tab:scaling}. The transport coefficient $f_3$ reaches its minima $f_3/T^2\approx -9.870$ at around $\mu/T\approx\pm 10.883$, and successively changes sign for large $|\mu|/T$. The transport coefficient $f_5$ reaches its symmetrically located extremas of $f_5/T\approx\pm 1.296$ at $\mu/T=\mp4.60$. At around $\mu/T\approx \pm10.883$, $f_5$ changes sign and starts to behave linearly in $\mu/T$ with $f_5/T\sim\pm2\pi + 0.4837 \mu/T$. The slope is very similar to the behavior at small $|\mu|/T$, even though its sign has now changed. It is interesting to note that $f_5/T$ changes sign for the exact same values of $\mu/T$ where the transport coefficient $f_3$ exhibits its minima. Note that at large $\mu/T$ the transport coefficient $f_7/T$ changes slope and behaves like $f_7/T\sim3.788\,\sign(\mu/T)+\mu/T$.

\begin{table}[h]
\begin{center}
\begin{tabular}{l c c c c c c c c c}
\toprule
$\alpha=2/\sqrt{3}\quad$& $\mu/T\to -\mu/T$	& Behaviour for $|\mu|/T\ll1$ 		\\ \addlinespace
\midrule
$f_1/T^2$		&+	&$-\frac{\pi^2}{2}-\frac16\, (\mu/T)^2$ \\
$f_2/T^2$		&+	&$-3\pi^2-(\mu/T)^2$  \\
$f_3/T^2$		&+	&$-\frac{\pi^2}{4}-0.2489\,(\mu/T)^2$  \\
$f_4$		&+	&$0.01230 \,(\mu/T)^2$ \\
$f_5/T$		&-	& $-0.4975 \,(\mu/T)$\\
$f_6$		&+	& $-0.3466-0.0266\,(\mu/T)^2$\\
$f_7/T$		&-	&$2\,(\mu/T)$ \\\bottomrule
\end{tabular}
\caption{Behaviour of the susceptibilities under a parity transformation $\mu\to-\mu$ and polynomial fits for small chemical potentials $|\mu|/T \ll 1$. Recall that the axial chemical potential $\mu$ is $\mathbb{P}$-odd and $\mathbb{C}$-even.}
\label{tab:scaling}
\end{center}
\end{table}

Note that in strong magnetic fields, where $B^\mu \sim \mathcal{O}(1)$ in the derivative counting scheme, the Kubo formula for $f_5$ is no longer second order but already has first order in derivative contributions. The first order Kubo formula for $f_5$ in the presence of a strong background magnetic field $B_0$ reads (compare with $M_4$ in \cite{Hernandez:2017mch} or $M_5$ in \cite{Ammon:2020rvg})
\begin{equation}
\label{eq:M2M5Kubo}
\begin{aligned}
\frac{1}{k_z} {\rm Im}\, G_{T^{tx} T^{yz}}(\omega = 0, k_z \mathbf{\hat{z}}) = - B_0 \, f_5 \,,
\end{aligned}
\end{equation}
and was computed in~\cite{Ammon:2020rvg}.
The authors found that
\begin{equation}\label{eq:M5Tch2}
\frac{\partial(f_5/T)}{\partial (\mu/T)}\Big|_{\mu/T=0} = - \frac{1}{2} + c_1\, B^2/T^4\, ,
\end{equation}
where $c_1$ is a negative constant. This is in perfect agreement with our results (to zeroth order in the magnetic field).

\subsection{Comments on vanishing anomaly $\alpha=0$}
As discussed in section \ref{holo_model}, the consistent truncation of type IIB supergravity leads to the Einstein-Maxwell-Chern-Simons action eq.\ \eqref{eq:bulk_action}, dual to the strongly coupled regime of the $\mathcal{N} = 4$ super-Yang-Mills theory with a $U(1)$ charge \cite{Chamblin:1999tk}. The parameter $\alpha$, which is the coefficient of the Chern-Simons term in the bulk action eq.\ \eqref{eq:bulk_action}, takes on the value $\frac{2}{\sqrt{3}}$. This is related to the strength of the chiral anomaly $C$ for the boundary super-Yang-Mills theory, $\nabla_\mu J^\mu = C E\cdot B$ , via $C = - \alpha = -\frac{2}{\sqrt{3}}$. There are no background electric or magnetic fields applied to the plasma in our setup, \textit{i.e.} $E, B = 0$, and thus one would not see the anomaly in the equation for current conservation. However, it will be manifest in other physical quantities, such as triangle diagrams involving the anomalous current, as well as the equations of motion eqs.\ \eqref{eq:eom1}, \eqref{eq:eom2}. In the results we have presented in this section, figure \ref{fig:plots}, the effects of this anomaly enter via the equations of motion satisfied by the bulk perturbations.

It is still instructive to consider the case when one sets the anomaly to vanish, $\alpha = 0$. The resulting bulk description in terms of the Einstein-Maxwell action is no more dual to the $\mathcal{N} = 4$ super-Yang-Mills plasma with a $U(1)$ charge on the $\textrm{AdS}_5$ boundary. It can rather be considered dual to the strongly coupled regime of some non-anomalous $U(1)$ charged conformal theory living on the boundary of $\text{AdS}_5$. It is important to note that the $U(1)$ symmetry in this case is vector in nature, and hence the charge conjugation and parity transformation properties will be different from the ones in table \ref{table:cpt}.  

When $\alpha \neq 0$, the Chern-Simons term couples the sectors with the fluctuations $\{a_x,\,h_{tx},\,h_{xz}\}$ and $\{a_y,\,h_{ty},\,h_{yz}\}$, while all other sectors stay unaffected by the chiral anomaly. Thus, from the Kubo formulas in table \ref{tableKubo}, it is clear that the transport coefficients $f_1,f_2,f_6$ and $f_7$ do not get influenced by the chiral anomaly, and remain the same for the non-anomalous case $\alpha=0$. However, the transport coefficients $f_3, f_4, f_5$ get affected, and we present their behaviour when $\alpha = 0$ in figure~\ref{fig:plots2} below. Note that the transport coefficient $f_4$ vanishes identically without the Chern-Simons term.
\begin{figure}[h!]
    \centering
  \includegraphics[width=0.35 \linewidth]{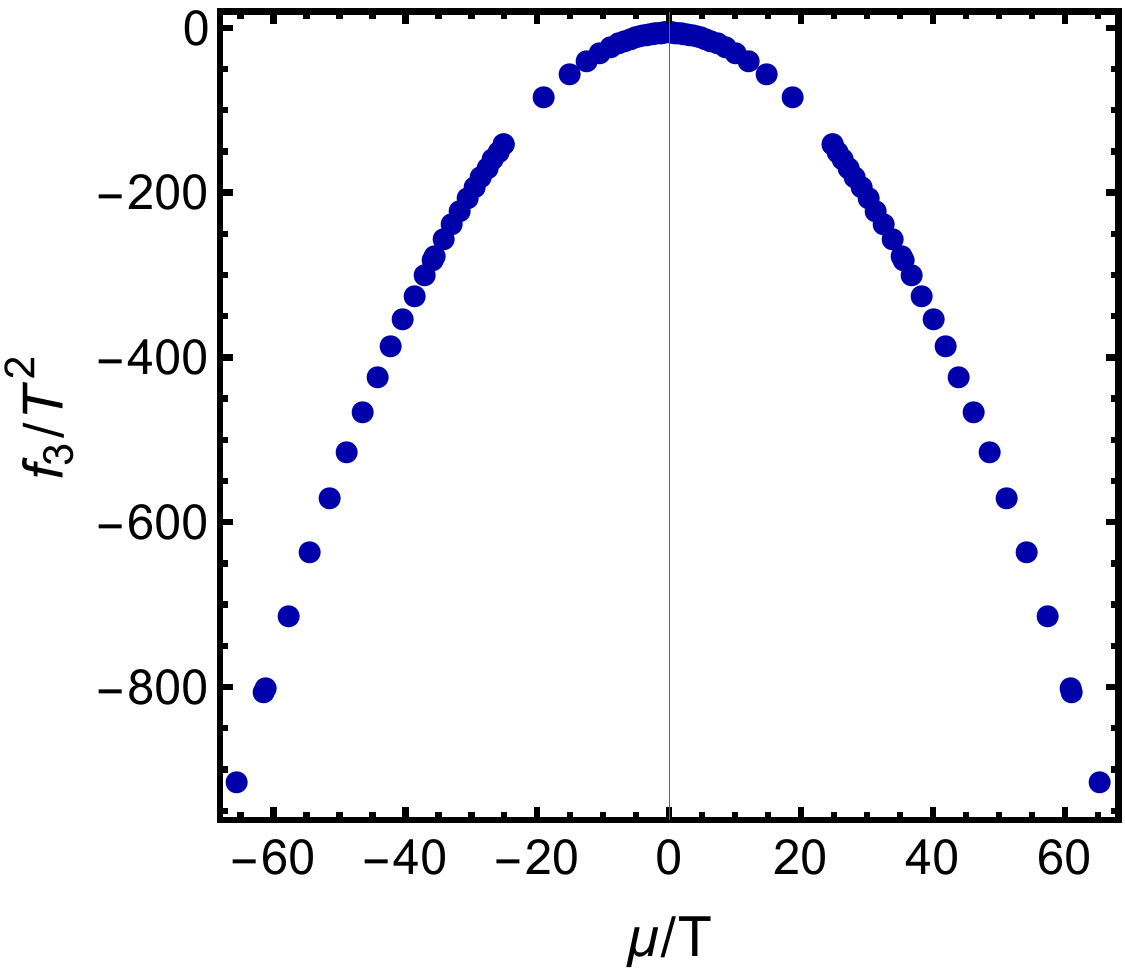} \qquad\qquad
  \includegraphics[width=0.34 \linewidth]{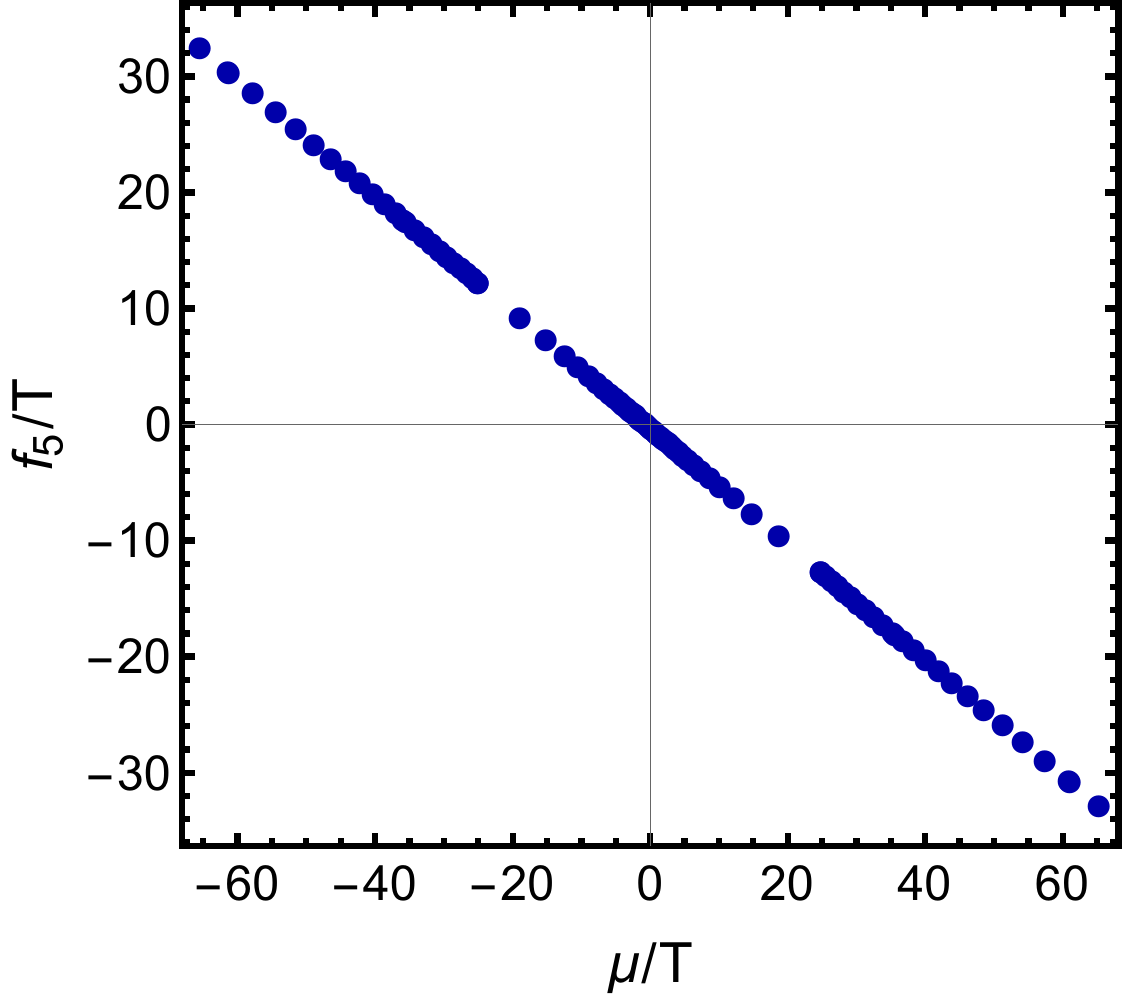}
    \caption{Behaviour of the dimensionless susceptibilities $f_3/T^2$, $f_5/T$ as functions of $\mu/T$ for the case of vanishing anomaly $\alpha = 0$. The susceptibility $f_4$ vanishes identically in this situation, while the remaining four susceptibilities $f_1, f_2, f_6$ and $f_7$ stay unaffected by setting $\alpha = 0$, with their behaviour still being given by the plots in fig.\ \ref{fig:plots}.}
    \label{fig:plots2}
\end{figure}

\begin{table}[h]
\begin{center}
\begin{tabular}{l c c c c c c c c c}
\toprule
$\alpha=0\quad$& $\mu/T\to -\mu/T$	& Behaviour for $|\mu|/T\ll1$ 		\\ \addlinespace
\midrule
$f_3/T^2$		&+	&$-\frac{\pi^2}{4}-\frac14\,(\mu/T)^2$  \\
$f_5/T$		&-	& $-\frac12\,(\mu/T)$ \\\bottomrule
\end{tabular}
\caption{Behaviour at $\alpha=0$ of the susceptibilities $f_3, f_5$ under a charge conjugation $\mu\to-\mu$ for the vector chemical potential, and polynomial fits for small dimensionless chemical potential $|\mu|/T \ll 1$.}
\label{tab:scaling2}
\end{center}
\end{table}

Similar to the case with the chiral anomaly, the second order transport coefficient $f_5$ may be computed in the presence of strong magnetic fields by the first order Kubo formula eq.~\eqref{eq:M2M5Kubo} when $\alpha = 0$.
The authors of \cite{Ammon:2020rvg,Bu:2019qmd} found that
\begin{equation}\label{eq:M5Tch}
\frac{\partial(f_5/T)}{\partial (\mu/T)}\Big|_{\mu/T=0} = - \frac{1}{2} + c_2\, B^2/T^4\, ,
\end{equation}
where $c_2$ is a positive constant. We again find perfect agreement with our results (to zeroth order in the magnetic field).

\subsection{Comments on the transport parameter $\kappa$ at $\mu = 0$}
In the uncharged case $\mu=0$, the susceptibility $f_1$ for the $\mathcal{N} = 4$ super-Yang-Mills plasma was computed in~\cite{Baier:2007ix, Bhattacharyya:2008jc}. In fact their result is quoted in terms of the second-order transport parameter $\kappa$ which is defined in the Landau-Lifshitz frame,\footnote{Note that $\kappa$ here should not be confused with the coefficient $\kappa_5$ appearing in the action eq.\ \eqref{eq:bulk_action} for the holographic model, i.e. $\kappa_5^2=8\pi G_N$.} and is related to our $f_1$ via a standard frame transformation (see appendix A of \cite{Kovtun:2018dvd}) through $\kappa = -2 f_1$. Their result
\begin{equation}
\label{eq:estimate}
   -2f_1= \kappa=\frac{\eta}{\pi\,T}=\frac{\eta}{s}\frac{s}{\pi T}=\frac{1}{4\pi}\frac{\frac{2\pi}{\kappa_5^2}\left(\frac{L}{u_h}\right)^3}{\pi T}=\frac{L^3}{2\,\kappa_5^2\,u_h^2},
\end{equation}
perfectly matches our analytical calculation of $f_1$ in eq.\ \eqref{eq:analyticalresult}.

The transport coefficient $\kappa$ was also computed using lattice gauge theory techniques for the $SU(N_c)$ Yang-Mills plasma, and it was found that for $N_c = 3$ \cite{Philipsen:2013nea}
\begin{equation}
    \kappa=0.36(15) T^2.
    \label{eq:lattice}
\end{equation}

We can also make a numerical estimate of $\kappa$ for QCD using our results. We first match the entropy density of our holographic model to that of lattice QCD as explained for e.g. in \cite{Ghosh:2021naw}. The Stefan-Boltzmann value of the entropy density of three flavor QCD is 
\begin{equation}
\label{eq:stefan}
  s_\text{SB}=4\,(\nu_b+7/4\,\nu_f)\,\frac{\pi^2\,T^3}{90}  \, ,
\end{equation}
with $\nu_b=2(N_c^2-1),\,\nu_f=2\,N_c\,N_f$ and $N_f=3=N_c$. Note that this value is only reached at asymptotically high temperatures. We can match this to the black hole entropy of our gravitational model at vanishing chemical potential,
\begin{equation}
\label{eq:bhentropy}
    s_\text{BH}=\frac{A}{4G_N\,\text{Vol}(\mathbb R^3)}=\frac{2\pi}{\kappa_5^2}\left(\frac{L}{u_h}\right)^3=\frac{4\pi^4\,L^3\,T^3}{2\kappa_5^2},
\end{equation}
where we have used eq.\ \eqref{eq:temp} in the last equality to express $u_h$ in terms of the temperature $T$.
If we want to compare eq.\ \eqref{eq:stefan} with $\mathcal N=4$ super-Yang-Mills plasma in the limit of infinite 't-Hooft coupling $\lambda$, eq.\ \eqref{eq:bhentropy}, we have to take into account a factor of 3/4 between the entropy densities \cite{Gubser:1996de,ammon_erdmenger_2015}, i.e. $3s_\text{SB}/4=s_\text{BH}$, yielding
\begin{equation}
    2\kappa_5^2=\frac{48\pi^2}{19}L^3.
\end{equation}
Using this result for $\kappa_5$ in eq.\ \eqref{eq:estimate} gives at $\mu=0$ an estimate of
\begin{equation}
   -2f_1=\kappa=\frac{L^3 \pi^2\,T^2}{2\,\kappa_5^2}=\frac{19}{48}\,T^2\approx 0.3958\,T^2.
\end{equation} 
 This is within the error range of the lattice result eq.\ \eqref{eq:lattice}. The difference is expected as the result of \cite{Philipsen:2013nea} is for pure Yang-Mills plasma, whereas in QCD one also has to take the quarks and their flavours into account.\footnote{For $N_f=0$, we find $2\kappa_5^2=15\, L^3\,\pi^2/2$ and thus $\kappa=2\,T^2/15\approx 0.1333 T^2$.} 
Additionally, the authors of \cite{Philipsen:2013nea} quote a much smaller value for $\kappa$ as the AdS/CFT result since they use a different entropy density too. Our computations thus provide an interesting comparison between holography and lattice gauge theory results.\footnote{Another estimate may be found in \cite{Romatschke:2009ng,Romatschke:2019gck} for an $SU(N_c)$ gauge theory,
$\kappa=(N_c^2-1)\, T^2/18=0.44 \,T^2$,
where we used $N_c=3$ in the last step. Furthermore, for 3 flavour QCD the authors in \cite{Moore:2012tc} found $\kappa=0.5694\,T^2.$}

\section{Discussion}
\label{discussion}
In this short article, we computed the seven $\mathbb{T}$-invariant second order thermodynamic susceptibilities $f_1, \ldots f_7$ for the charged strongly coupled $\mathcal{N} = 4$ super-Yang-Mills plasma and illustrated their dependence on the dimensionless parameter $\mu/T$ (see fig.\ \ref{fig:plots}). To compute the susceptibilities, we employed Kubo formulas that gave these susceptibilities in terms of equilibrium two-point functions of the energy-momentum tensor and the $U(1)$ current. We computed the two-point functions holographically, by studying fluctuations to the asymptotically AdS$_5$ Reissner-Nordstr\"{o}m black brane geometry. As also mentioned in the Introduction, studying the transport properties of the $\mathcal{N} = 4$ super-Yang-Mills plasma via holography provides insights into the expected behaviour of the quark-gluon plasma produced in heavy ion collisions. 

Interestingly, we also computed the second order transport coefficient $\kappa$ analytically, which is related to the susceptibility $f_1$, and measures the response of the fluid to the presence of background curvature. Using the analytic result, an estimate could be made for the value of $\kappa$ for QCD. Our calculations give a result in proximity of the lattice gauge theory computations for Yang-Mills plasma, giving further credibility to the idea that transport properties of strongly coupled QCD matter can be captured via holographic techniques modeling the super-Yang-Mills plasma.

A few comments are in order. The super-Yang-Mills theory is a conformal field theory, and has a trace anomaly in the presence of external electromagnetic fields, given by\footnote{There are further contributions to the trace anomaly if the conformal theory is put on a curved background, proportional to $R^2$. However, these are fourth order in derivatives and irrelevant for the present discussion.}
\begin{equation}
    \langle T^\mu_\mu \rangle = f'_4 B^2 + f'_6 E^2.
\end{equation}
Here we have used the notation $f'_n = T \partial f_n / \partial T + \mu \partial f_n / \partial \mu$. For us, $f_4, f_6 \neq 0$. However, since we apply no external electromagnetic fields to the super-Yang-Mills plasma \textit{i.e.} $E = B = 0$, the trace anomaly must vanish. That this indeed happens is very easy to see by computing the stress tensor components and evaluating the trace, since we know the background solution analytically. 

We computed the behaviour of the second order susceptibilities as a function of $\mu/T$ at strong coupling via holographic techniques. At the other extreme, when the plasma can be considered weakly coupled, it is much simpler to compute the same susceptibilities. The matter content of the $\mathcal{N} = 4$ super-Yang-Mills theory consists of four Weyl fermions, six real scalars and a vector field, all transforming in the adjoint representation of $SU(N_c)$. These matter fields can be approximately considered as mutually non-interacting in the limit of very weak coupling. Now the contribution to the susceptibilities from each of these individual matter fields was already computed in \cite{Kovtun:2018dvd}. For instance, in the uncharged case, for a real scalar one has
\begin{equation}
    f_1 = \frac{T^2}{144}(1-6\xi)\, , \quad f_2 = 0\, , \quad f_3 = - \frac{T^2}{144} \, .
\end{equation}
Here $\xi$ denotes the coupling of the scalar to the background curvature,  with the field being minimally coupled for $\xi = 0$ and conformally coupled for $\xi = 1/6$. Similarly, the susceptibilities for a Weyl fermion are
\begin{equation}
    f_1 = - \frac{T^2}{288}\, , \quad f_2 = - \frac{T^2}{48} \, , \quad f_3 = - \frac{T^2}{576} \,.
\end{equation}
For the adjoint vector field we have
\begin{equation}
    f_1 = - (N_c^2 -1)\, \frac{T^2}{36}\, , \quad f_2 = - (N_c^2 -1)\, \frac{T^2}{6} \, , \quad f_3 =  (N_c^2 -1) \, \frac{T^2}{36} \, .
\end{equation}
In the weakly coupled limit, the net result for the susceptibilities of the uncharged $\mathcal{N} = 4$ super-Yang-Mills plasma can therefore easily be read off from the results above by simply adding the contributions coming from each of the matter fields appropriately. One must keep in mind that since the scalar field as well as the Weyl fermion transform in the adjoint representation of the gauge group, the total degrees of freedom associated with each scalar and Weyl fermion is $N_c^2 -1$ \cite{Yamada:2006rx}. With this counting taken into account, it is easy to see that for the super-Yang-Mills plasma at weak coupling we have
\begin{equation}
    f_1 = - (N_c^2 -1)\frac{T^2}{24}\, , \quad f_2 = - (N_c^2 -1)\frac{T^2}{4}\, , \quad f_3 = - (N_c^2 -1)\frac{T^2}{48}\, .
\end{equation}
In the limit of large $N_c$, this implies that for the weakly coupled plasma
\begin{equation}
    f_1 = \frac{1}{6} f_2 = 2 f_3 \approx - \frac{N_c^2 T^2}{24}.
\end{equation}
Now, in the strongly coupled regime, the uncharged super-Yang-Mills plasma has
\begin{equation}
    f_1 = \frac{1}{6} f_2 = 2 f_3 = - \frac{\pi^2 T^2}{2}\, ,
\end{equation}
as can be easily read off from table \ref{tab:scaling}. The fact that the ratio $f_1/f_2$ stays the same in going from strong to weak coupling is an artefact of the theory being conformal: it is the Weyl invariance of the generating functional that enforces this constraint \cite{Kovtun:2018dvd}. However, $f_3$ is not constrained by conformality, and therefore it comes out as a surprise that the ratio $f_1/f_3$ (or $f_2/f_3$) remains the same, irrespective of whether one is in the strong or the weak coupling regime. 

Let us conclude the article by pointing out that determining the second order equilibrium susceptibilities for the $\mathcal{N} = 4$ super-Yang-Mills theory provides a powerful though indirect step towards an understanding of the second order equilibrium transport properties of the quark-gluon plasma. The quark-gluon plasma is produced at heavy ion colliders such as the LHC and RHIC in an out-of-equilibrium state due to thermal fluctuations. Needless to say, there is a very rich spectrum of physical effects that arise as a consequence of these out of equilibrium fluctuations, and the second order equilibrium transport properties discussed here will receive important corrections due to the fluctuation effects. Large magnetic fields are also produced during heavy ion collisions \cite{Kharzeev:2007jp, Skokov:2009qp, Bzdak:2011yy} making the background anisotropic, and their presence can also significantly affect the transport and relaxation phenomena occurring in the quark-gluon plasma \cite{Kharzeev:2012ph, Ammon:2020rvg, Shukla:2021ksb,Ghosh:2021naw}. Also, the remaining two second order susceptibilities - $f_8, f_9$ - which are $\mathbb{T}$-odd and were not the subject of study in the current paper, can in principle be determined using equilibrium three-point functions of the energy-momentum tensor and the current \cite{Kovtun:2018dvd}, although closed form expressions of Kubo formulas for them are not yet known.\footnote{The susceptibility $f_9$ vanishes for conformal theories, which is the case for the $\mathcal{N} = 4$ super-Yang-Mills plasma.} All these possible directions to extend the present work are extremely interesting - we leave them out here for near future investigations.

\acknowledgments
We would like to thank Martin Ammon, Pavel Kovtun and Karl Landsteiner for helpful discussions and comments. SG is supported by the `Atracci\'on de Talento' program (2017-T1/TIC-5258, Comunidad de Madrid) and through the grants SEV-2016-0597 and PGC2018-095976-B-C21.
Research at Perimeter Institute is supported in part by the Government of Canada through the Department of Innovation, Science and Economic Development Canada and by the Province of Ontario through the Ministry of Colleges and Universities.

\appendix

\section{Analytic computation of $f_1$}
\label{app:analytical}
In this appendix, we present an analytic calculation to obtain the two-point function $G_{T^{xy}T^{xy}}$, which can be used for determining the susceptibility $f_1$ via the Kubo formula
\begin{equation}
    f_1=-\frac 12\lim\limits_{{\bf k}\to 0}\frac{\partial^2}{\partial k_z^2} G_{T^{xy}T^{xy}}\, ,
\end{equation}
following the techniques laid out in \cite{Ammon:2020rvg,Grieninger:2020wsb}.
The relevant equation of motion for the $h_{xy}$ fluctuations about the background Reissner-Nordstr\"{o}m black brane solution decouples,
\begin{equation}
\begin{split}
   3 g^2 k^2 L^2 u_h^4\, u\, h_{xy}(u) &+ \left(3 g^2 L^2 \left(3 u_h^4+u^4\right)+2 \kappa_5 ^2 \mu ^2 u^4 \left(u_h^2-3 u^2\right)\right)h_{xy}'(u) \\&-u \left(u_h^2-u^2\right)  \left(3 g^2 L^2 \left(u_h^2+u^2\right)-2 \kappa_5 ^2 \mu ^2 u^4\right)h_{xy}''(u)=0,
\end{split}
\end{equation}
where the asymptotic expansion of $h_{xy}$ is given by
\begin{equation}
    h_{xy}=h^{(0)}_{xy}-\frac{1}{4}h^{(0)}_{xy} k^2\,u^2+\frac{1}{24}\left(h^{(4)}_{xy}-\frac 32\,h^{(0)}_{xy} k^4\,\log(u)\right)+\mathcal O(u^6)\, .\label{eq:asympexp}
\end{equation}
Note that the second order Green's function is unbothered by the logarithmic term scaling with $k^4$.
We can compute the Green's function $G_{T^{xy} T^{xy}}$ analytically by expanding the fluctuation $h_{xy}$ to second order in $k$, i.e. \begin{equation}
    h_{xy}=h_{xy}^{(0)}+h_{xy}^{(1)}\,k+h_{xy}^{(2)}\,k^2\, ,
\end{equation}
and solving the equation of motion order by order in $k$, with the boundary condition $h_{xy}^{(0)}(0)=c_1, \ h_{xy}^{(1)}(0)=0,\ h_{xy}^{(2)}(0)=0$. With this, the only solution regular at the horizon is
\begin{align}
   & h_{xy}^{(0)}(u)=c_1\, ,\quad  h_{xy}^{(1)}(u)=0\, ,\\&
   h_{xy}^{(2)}(u)=\frac{c_1\, g\, L\, u_h^2}{2 \sqrt{g^2 L^2+\frac{8}{3} \kappa_5 ^2 \mu ^2 u_h^2}} \left[\!\arccoth\!\left(\!\frac{g L \sqrt{9 g^2 L^2\!+\!24 \kappa_5 ^2 \mu ^2 u_h^2}}{3 g^2 L^2-4 \kappa_5 ^2 \mu ^2 u^2}\right)\!\!-\!\arctanh\! \left(\frac{g L}{\sqrt{g^2 L^2+\frac{8}{3} \kappa_5 ^2 \mu ^2 u_h^2}}\right)\right].
\end{align}
Expanding this solution at the boundary $u=0$ yields
\begin{equation}
    h_{xy}^{(2)}=-\frac{c_1}{4} u^2+\frac{c_1}{8\,u_h^2} u^4+\mathcal O(u^6),
\end{equation}
which perfectly matches the asymptotic expansion in eq.\ \eqref{eq:asympexp}. 
For $c_1=1$ we find for the expectation value 
\begin{equation}
f_1=-\frac 12\lim\limits_{{\bf k}\to 0}\frac{\partial^2}{\partial k_z^2} G_{T^{xy}T^{xy}}=-\frac 12\cdot\frac{1}{6}\cdot \frac{3\,L^3}{2\kappa_5^2\,u_h^2}\cdot 2=-\frac{L^3}{4\kappa_5^2\,u_h^2}\, .\label{eq:analyticalresult}
\end{equation}
The explicit dependence of $f_1$ on $\mu$ and $T$ can now easily be obtained by eliminating $u_h$ from the expression above using eq.\ \eqref{eq:temp}. This leads to the following result for $f_1$,
\begin{equation}
    f_1 = - \, \frac{\mu^4  \kappa_5^2 L^3}{\left(\sqrt{9\pi^2 T^2 g^4 L^4 + 12 \mu^2 g^2 \kappa_5^2 L^2} - 3\pi T g^2 L^2\right)^2}.
    \label{eq:f1anal2}
\end{equation}
This result is presented in eq.\ \eqref{eq:f1anal} of the main text, in the simplified set of units $g^2 = 2\kappa_5^2 = L =1$.

\section{Numerical methods}
\label{app:numericalmethods}
We determine the two-point functions required to compute the second order thermodynamic transport coefficients numerically by means of a pseudo-spectral method. In this appendix, we briefly explain the numerical techniques used; for a detailed introduction see for example~\cite{Boyd00}. Our discussion follows \cite{Ammon:2016fru,Grieninger:2020wsb,Grieninger:2017jxz,Baggioli:2019abx}. The basic idea of spectral methods is to account for the radial dependence of the unknown functions in the ordinary differential equations governing the fluctuations by expanding them in terms of a (truncated) series in the basis functions. The basis functions are analytically known and so are their radial derivatives. To set up the spectral method, we choose a discrete, finite collocation grid, a set of basis functions $\{T_k\}$ and a finite number $N$ at which we truncate the series in the basis functions, \begin{equation}
     X(u_i)=\sum_{k=0}^\infty c_i\,T_k(u_i)\approx X_N(u)=\sum\limits_{k=0}^{N-1}\,c_k\,T_k(u).\label{eq:spectralrep}
 \end{equation} 
Spectral methods are global methods, which means that we take the function values on the whole domain into account when we compute numerical derivatives. This is in strong contrast to finite difference methods, which require only the values from the neighboring grid-points.

We discretize the equations of motion by a Chebyshev-Lobatto grid with $N$ grid-points,
\begin{equation}
    u_m=\frac12\left(\cos\left(\frac{m\,\pi}{N-1}\right)+1\right), \quad m\in [0,N-1].\label{eq:chebylobatto}
\end{equation} 
 In this work, we choose the Chebyshev polynomials $T_k=\cos(k\,\arccos(2u-1))$, where $u\in[0,1]$ as basis functions. 
The derivatives acting on the functions simply translate into derivatives of the basis functions, 
\begin{align}
    &X'(u)\, \approx\sum\limits_{n=0}^{N-1} c_n\,T_n'(u)\,=\sum\limits_{n,m=0}^{N-1}c_n\,\hat D_{nm}T_m(u)=\sum\limits_{n=0}^{N-1}c_n'\,T_n(u),\\
    &X''(u)\approx\sum\limits_{n=0}^{N-1} c_n\,T_n''(u)=\sum\limits_{n,m,l=0}^{N-1}c_n\,\hat D_{nm}\hat D_{ml}\,T_l(u)=\sum\limits_{n=0}^{N-1}c_n''\,\phi_n(u).
\end{align}
With the derivative matrix $\hat{D}$, we can compute the discrete derivatives on our grid. Since spectral methods are global methods (in contrast to finite differences), the information on each grid-point is needed for the derivative (instead of just the neighborhood). In the context of holography, this is very convenient since it allows us to impose boundary conditions on both ends of the domain simultaneously.

We can determine the coefficients of the derivatives simply from the original coefficients by multiplying the original coefficients with a derivative matrix.
In this work, we do not work directly with the coefficients but with the values of the functions on the discretized grid (this is referred to as pseudo-spectral method). We can translate from coefficients to the values of the functions with
\begin{equation}
    X_i= \sum_{k=0}^{N-1}c_k T_k(u_i), 
\end{equation}
which gives us the value of the function X at $u_i$.

\begin{figure}[h!]
    \centering
    \includegraphics[width=0.48 \linewidth]{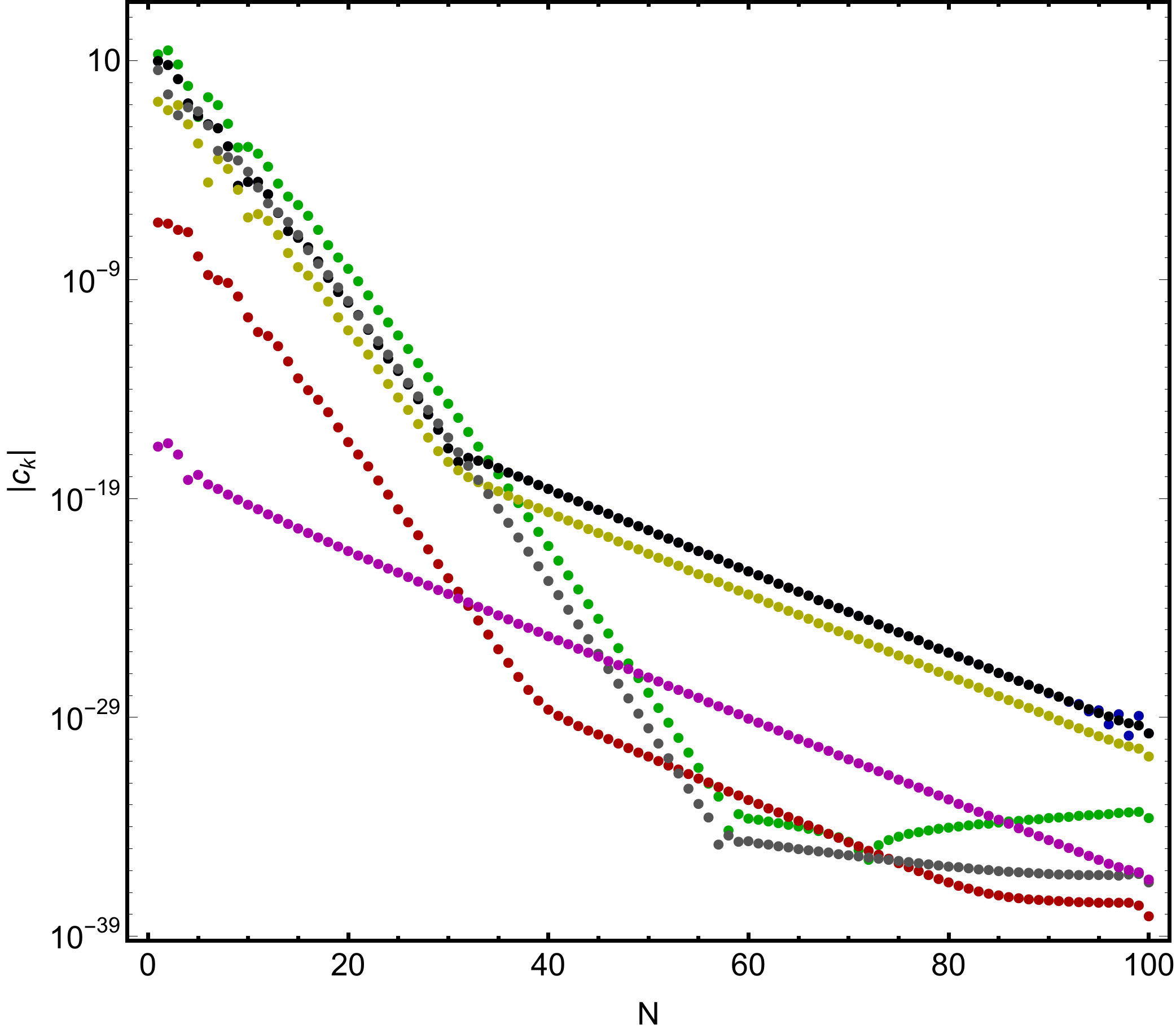}\quad \includegraphics[width=0.48 \linewidth]{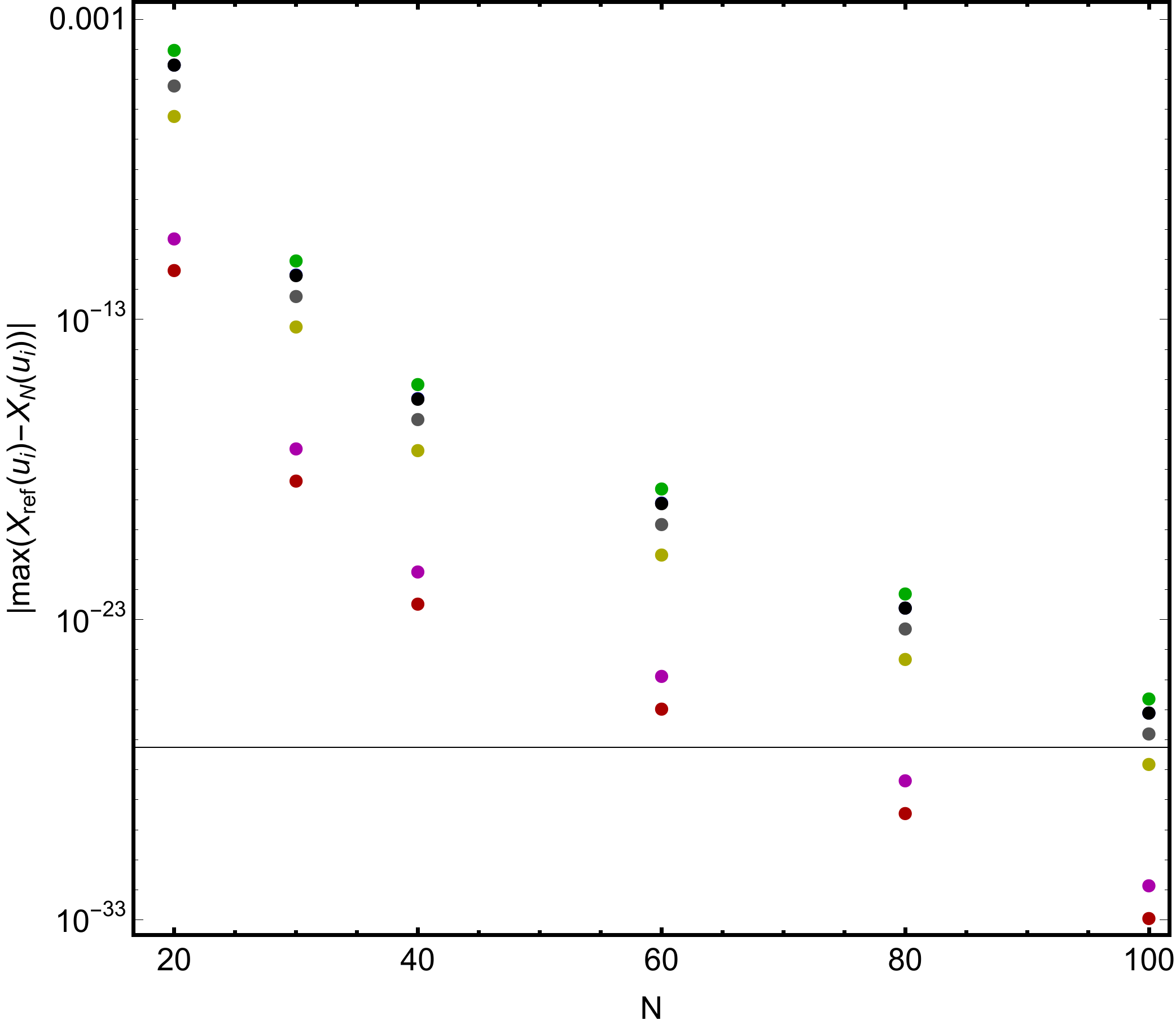}
    \caption{The fluctuations $h_{tt},\,h_{tz},\,h_{xx},\,h_{yy},\,h_{zz},\,a_t,\,a_z=$ (green, red, blue, black, yellow, gray, magenta) for $h_{tt}(0)=1,\,\mu/T=61.062,\,k/T=0.0002655$. \textbf{Left:} Chebyshev coefficients of the numerical solution. \textbf{Right:} Maximal deviation from the numerical solution with $N=130$ grid-points evaluated on an equidistant grid as function of the number of grid-points.}
    \label{fig:convergence}
\end{figure}

The recipe for obtaining the numerical solution goes as follows:
\begin{enumerate}
\item Choose a number of grid-points $N$ and discretize the equations of motion on the Chebyshev-Lobatto grid eq.\ \eqref{eq:chebylobatto}, and replace all derivatives by acting with the derivative matrices on the discretized functions, i.e. $X_i'=\sum_{m=0}^{N-1}\hat D_{im} X_m$. This results in a $M\,N\times M\,N$ dimensional matrix $\hat K$, where $M$ denotes the number of unknown functions in our system of differential equations.
\item Boundary conditions, if not already introduced as behavioral boundary conditions, can be imposed by replacing the corresponding $1+(i-1)\,N$ and $i\,N$th row, where $i$ is the $i$th function. 
\item Solve the corresponding linear system
\begin{equation}
    \hat K \bm X=\phi,
\end{equation}
where $\hat K$ is the discretized matrix (see step 1 above), $\bm X$ a $M\, N$ dimensional vector with the discretized functions and $\phi$ the right hand side with the values independent of the functions.
\item Check the convergence of the numerical solution.
\end{enumerate}
To confirm the accuracy of our numerics, we perform several checks. First of all, we check that the Chebyshev coefficients of the numerical solution drop sufficiently fast and below our threshold, as presented in the left plot of figure \ref{fig:convergence}. Furthermore, we check that the equations of motion and constraint equations are satisfied on an equidistant grid. Finally, we show that the functions converge to an accurate solution (right side of figure \ref{fig:convergence}). For the accurate solution we choose $N=130$ since the change of the solution for taking even more grid-points is smaller than $10^{-40}$.

To show that the numerical solutions converge to the accurate solution, we compute numerical solutions with less grid-points. We compare these ``less accurate" solutions to the solution with $N=130$ on an equidistant grid with $z=i/129$ for $i=0,\ldots,129$ and compute the biggest deviation. We can see that solutions converge exponentially towards the $N=130$ solution.

\bibliography{main}

\end{document}